\documentclass[pdflatex,sn-nature]{sn-jnl}

\usepackage{graphicx}%
\usepackage{multirow}%
\usepackage{amsmath,amssymb,amsfonts}%
\usepackage{amsthm}%
\usepackage{mathrsfs}%
\usepackage[title]{appendix}%
\usepackage{xcolor}%
\usepackage{textcomp}%
\usepackage{manyfoot}%
\usepackage{booktabs}%
\usepackage{algorithm}%
\usepackage{algorithmicx}%
\usepackage[left]{lineno}  
\usepackage{caption}%
\usepackage{algpseudocode}%
\usepackage{listings}%

\theoremstyle{thmstyleone}%

\theoremstyle{thmstyletwo}%

\theoremstyle{thmstylethree}%

\begin{document}

\title[Article Title]{Solving the inverse problem of X-ray absorption spectroscopy via physics-informed deep learning}

\author[1]{\fnm{Suyang} \sur{Zhong}}
\equalcont{These authors contributed equally to this work.}
\author[1]{\fnm{Boying} \sur{Huang}}
\equalcont{These authors contributed equally to this work.}
\author[2]{\fnm{Pengwei} \sur{Xu}}
\equalcont{These authors contributed equally to this work.}
\author[2]{\fnm{Fanjie} \sur{Xu}}
\author[2]{\fnm{Yuhao} \sur{Zhao}}

\author*[3,2,4]{\fnm{Jun} \sur{Cheng}}\email{chengjun@xmu.edu.cn}
\author*[1,2,4]{\fnm{Fujie} \sur{Tang}}\email{tangfujie@xmu.edu.cn}

\author[5,6,7]{\fnm{Weinan} \sur{E}}
\author[3,4]{\fnm{Zhong-Qun} \sur{Tian}}

\affil[1]{\orgdiv{Pen-Tung Sah Institute of Micro-Nano Science and Technology, Discipline of Intelligent Instrument and Equipment, iChEM}, \orgname{Xiamen University}, \orgaddress{\city{Xiamen}, \postcode{361005}, \country{China}}}

\affil[2]{\orgdiv{Institute of Artificial Intelligence}, \orgname{Xiamen University}, \orgaddress{\city{Xiamen}, \postcode{361005}, \country{China}}}

\affil[3]{\orgdiv{State Key Laboratory of Physical Chemistry of Solid Surfaces, iChEM, College of Chemistry and Chemical Engineering}, \orgname{Xiamen University}, \orgaddress{\city{Xiamen}, \postcode{361005}, \country{China}}}

\affil[4]{\orgdiv{Laboratory of AI for Electrochemistry (AI4EC), Tan Kah Kee Innovation Laboratory (IKKEM)}, \orgaddress{\city{Xiamen}, \postcode{361005}, \country{China}}}

\affil[5]{\orgname{AI for Science Institute}, \orgaddress{\city{Beijing}, \postcode{100080}, \country{China}}}

\affil[6]{\orgdiv{Center for Machine Learning Research}, \orgname{Peking University}, \orgaddress{ \city{Beijing}, \postcode{100871}, \country{China}}}

\affil[7]{\orgdiv{School of Mathematical Sciences}, \orgname{Peking University}, \orgaddress{ \city{Beijing}, \postcode{100871}, \country{China}}}
    \abstract{Resolving transient atomic configurations in non-crystalline or dynamic environments remains a fundamental bottleneck in the physical sciences. While X-ray absorption spectroscopy (XAS) is a premier probe of local structure, inverting spectra into structural descriptors is a notoriously ill-posed problem due to inherent many-to-one mapping. Here, we present the Spectral Pattern Translator (SPT), a physics-informed deep learning framework that establishes a robust bridge between large-scale theoretical datasets and experimental reality. Our strategy exploits the Fourier duality between spectral energy oscillations and spatial scattering paths to overcome the  ``simulation-to-experiment" gap. By decomposing spectra into frequency domains, SPT effectively isolates robust structural coordination signals from the destabilizing noise inherent in experimental data. Trained on a massive library of diverse atomic environments, this approach achieves state-of-the-art accuracy in resolving continuous phase transitions in battery cathodes and deciphering local order in amorphous materials. With millisecond-scale latency, SPT removes the primary computational barrier to autonomous materials discovery, establishing a robust, noise-resilient engine for closed-loop robotic chemistry.}
    
    \keywords{X-ray Absorption Spectroscopy, First-principles Calculations, Structure-Spectrum Relationships, Deep Learning, Computational Materials Science}

\maketitle

\section{Introduction}\label{sec1}

    \par The ability to resolve the precise atomic arrangements and electronic transitions of matter in real-time, especially within disordered or non-crystalline regimes, remains one of the fundamental frontiers of modern physical science. While crystalline structures can be mapped with high precision, the ``dark matter" of materials science, from amorphous catalysts~\cite{smith2013photochemical,kang2023cr}, dynamic battery intermediates~\cite{liu2021dynamic}, and transient states~\cite{ostrom2015probing,van2017situ}, requires tools that can probe local coordination without the need for long-range order. X-ray absorption spectroscopy (XAS) is a powerful characterization technique for this purpose, widely used across physics, chemistry, and materials science~\cite{Lomachenko2016, Martini2017, Timoshenko2021}. In an XAS experiment, core electrons are excited by incident X-ray photons into unoccupied electronic states of the sample, and the resulting photoelectrons emitted from the absorbing atoms are scattered by neighboring atoms~\cite{zhu2021emerging, virga2023structural}. Consequently, XAS and in particular its near-edge region, known as X-ray absorption near-edge structure (XANES), encodes rich information about the local chemical environment of the absorbing site, such as oxidation state, coordination number, and local structural symmetry. Thus, XAS provides essential element-specific and site-sensitive insights~\cite{ravel2010simultaneous,farges1997ti,kerr2022characterization} into the structural and electronic properties of materials, which are critical for understanding their underlying physical and chemical processes.
    
    \par However, the quantitative analysis of XANES remains technically challenging due to the complex correlation between local atomic structures and their corresponding XAS features during physical and chemical processes~\cite{Nikolay2025, Guo2021, Liu2014}. 
    While the ``forward problem", calculating a spectrum from a known structure, is well-defined by quantum mechanics~\cite{rehr2000theoretical}, the ``inverse problem", inferring the atomic structure $\rho(r)$ from an observed spectrum $\mu(E)$, is notoriously difficult. Fundamentally, the task of inferring structural descriptors from spectral data constitutes a classic inverse problem. Mathematically, the relationship between the observed spectral function $\mu(E)$ and the underlying local structure distrubution $\rho(\mathbf{r})$ can be conceptualized as a non-linear Fredholm integral equation~\cite{arveson2001short}: 
    \begin{equation}
     \label{eq:Fredholm}
        \mu(E) = \int_{\Omega} \mathcal{K}(E, \mathbf{r}) \rho(\mathbf{r}) \, d\mathbf{r} + \epsilon,
    \end{equation}
    where $\mathcal{K}(E, \mathbf{r})$ represents the complex, energy-dependent scattering kernel governed by Fermi's golden rule and multiple scattering theory, and $\epsilon$ denotes experimental noise. This formulation is notoriously ill-posed, as the forward operator represented by the kernel $\mathcal{K}$ is compact, which implies that its inverse is unbounded. Consequently, the mapping from the low-dimensional spectral domain back to the high-dimensional structural domain is unstable: trivial perturbations in the input spectrum (noise $\epsilon$) can amplify into arbitrarily large errors in the reconstructed structural descriptors $\rho(\mathbf{r})$. This instability is exacerbated by the non-uniqueness of the solution, where distinct structural configurations may yield indistinguishable spectral signatures due to the loss of high-frequency spatial information during the scattering process~\cite{suenaga2010atom,aucour2023coupling}. Strictly speaking, recovering the full high-dimensional coordinate distribution $\rho(r)$ from the low-dimensional spectral signal $\mu(E)$ is mathematically ill-posed and often physically undefined for dynamic or disordered systems. However, we argue that the substantive inverse problem in materials science is the retrieval of statistical descriptors (e.g., coordination numbers, mean bond lengths), which represent the physically effective variables (moments) of $\rho(r)$. Unlike the unstable coordinate reconstruction, the inference of these descriptors can be formulated as a well-posed parameter estimation problem, provided the model incorporates sufficient physical constraints to regularize the solution. 

    \par Current approaches struggle to overcome these barriers, creating a bottleneck for the emerging era of Autonomous Materials Discovery~\cite{Burger2020roboticchemist, Zhang2025AI, Song2025AIChemist}. Empirical fingerprinting relies on matching spectra to limited libraries of known reference standards~\cite{ravel2005athena,protsenko2025fingerprint}, a method that fails for novel, amorphous, or dynamic systems where no such standards exist~\cite{zhu2021emerging, virga2023structural, kwon2024spectroscopy}. Conversely, $ab$ $initio$ simulations (such as core-hole potental, TD-DFT, GW-BSE, or real-space multiple scattering)~\cite{rehr2000theoretical, Galli2006, Vinson2011, Rehr2021, tangpnas2022} provide a theoretical basis for analysis but incur computational costs that scale cubically or exponentially with system size. This computational latency renders traditional methods incompatible with closed-loop ``robotic chemists"~\cite{Burger2020roboticchemist, Zhang2025AI, Song2025AIChemist}, which require real-time structural feedback to navigate vast chemical spaces autonomously. Within this context, ML has been increasingly employed to establish quantitative relationships between spectra and material properties, encompassing both forward modeling, predicting spectra from atomic structures~\cite{rankine2022accurate,Deyu2025omnixas} and inverse modeling, inferring structural parameters from spectral features~\cite{chen2024robust,na2025interpretable}. Early ML models, such as Random Forests, Extra Trees~\cite{Torrisi2020, wang2023interpretable,gleason2024prediction}, relied on manually crafted spectral descriptors for regression or classification tasks. The advent of deep learning brought end-to-end models that learn structural feature representations directly from raw or pre-processed spectral data~\cite{rankine2022accurate,wang2023interpretable,penfold2024machine}, eliminating manual feature design and achieving superior performance in identifying local chemical environments during the physical and chemical processes. Graph neural networks (GNNs)~\cite{zhan2025graph} represent an important step forward, capturing intra spectral correlations and structure–spectrum relationships with improved fidelity. Furthermore, existing machine learning attempts~\cite{carbone2019classification,martini2020pyfitit, rankine2022accurate, Deyu2023Carbon, Deyu2020QM9, Deyu2025omnixas, Deyu2023AAE, zhan2025graph} often lack the generalization capability required to serve as a foundational framework for the Genesis Mission, failing to transfer knowledge between crystalline and amorphous regimes.

   \par Here we present the Spectral Pattern Translator (SPT), a physics-informed deep learning framework that solves this inverse problem by introducing frequency-domain regularization as a strong inductive bias. We posit that transforming the spectral signal into the frequency domain effectively ``diagonalizes" the interaction operator, segregating robust physical signals from destabilizing noise. By exploiting the Fourier duality between energy-space oscillations ($k$-space) and radial scattering paths ($R$-space), the SPT isolates low-frequency components corresponding to dominant short-range interactions from high-frequency components that encode disorder or noise. Trained on a massive dataset of over 50,000 computed spectra, SPT achieves state-of-the-art accuracy with millisecond-scale inference speeds suitable for real-time integration into autonomous laboratories~\cite{Burger2020roboticchemist, Zhang2025AI, Song2025AIChemist}. This architecture establishes a robust, generalized paradigm for spectroscopic inversion, bridging the gap between theoretical simulation and experimental reality.

\section{Results}

\subsection{A Physics-informed Frequency-domain Framework}

    \begin{figure*}[t]
    \centering
    \includegraphics[width=\linewidth]{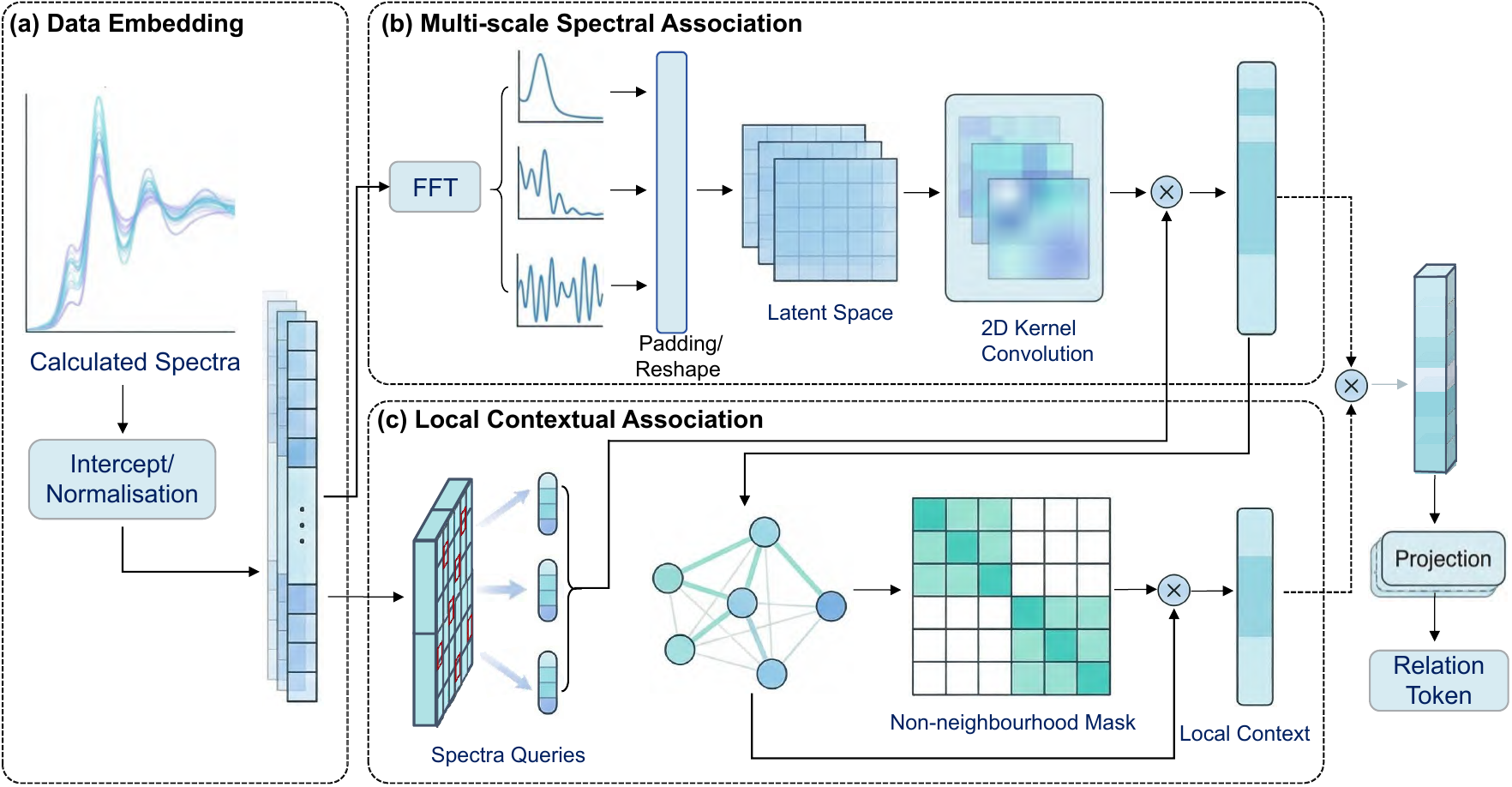}
        \caption{
            \textbf{Architecture of the Spectral Pattern Translator (SPT).}
            (a) Raw spectral data is preprocessed to extract key feature points, normalized, and encoded into initial spectral sequence vectors.
             (b) Dominant frequencies are identified via Fast Fourier Transform (FFT) on the spectral sequence. These components are reshaped into a two-dimensional tensor and fused through a learnable spectral query mechanism.
            (c) Feature nodes are embedded into a spatial graph structure, establishing dynamic contextual associations with local neighbors to refine structural representations.
        }
    \label{fig:model}
    \end{figure*}

    \par To resolve the non-uniqueness and instability inherent in the Fredholm integral equation of XAS (Eq.~\ref{eq:Fredholm}), we developed the SPT model, a deep learning framework designed to invert raw XANES spectra into probabilistic structural descriptors. Unlike conventional ``black-box" neural networks that operate solely on pixel-level correlations, the SPT introduces a strong inductive bias based on the physics of photoelectron scattering: frequency-domain regularization. The architecture (Figure~\ref{fig:model}a) processes spectral data through two parallel but interacting pathways: a \textbf{Multi-scale Spectral Association} module operating in the frequency domain, and a \textbf{Local Contextual Association} module operating in the graph domain.
    
    \par The core lies in the frequency-domain decomposition (Figure~\ref{fig:model}b). In X-ray absorption theory (specifically EXAFS), spectral oscillations in energy space ($k$-space) are mathematically conjugate to the radial distribution of scattering atoms in real space ($R$-space) via a Fourier transform. We exploit this duality by applying a Fast Fourier Transform (FFT) to the input XANES spectrum, effectively ``diagonalizing" the scattering interaction operator. This transformation segregates the signal based on scattering path lengths:
       \begin{itemize}
           \item \textbf{Low-frequency components}: encode the dominant, short-range interactions, such as the coordination number (CN) of the first shell and oxidation state (OS).
           \item \textbf{High-frequency components}: encode long-range disorder or, critically, experimental noise ($\epsilon$).
       \end{itemize}
    By projecting the data into this latent frequency space, the SPT utilizes 2D convolution layers as learnable band-pass filters. These filters are trained to selectively amplify the low-to-medium frequency bands that contain robust structural information while suppressing the high-frequency noise that typically destabilizes inverse solutions. This mechanism effectively converts the ill-posed global inversion problem into a series of well-posed, scale-specific feature extraction tasks.
    
    \par While frequency analysis captures global scattering modes, subtle electronic effects, such as pre-edge features arising from $d$-$p$ orbital hybridization or Jahn-Teller distortions, manifest as localized correlations between specific energy points. To capture these, the \textbf{Local Contextual Association} module (Figure~\ref{fig:model}c) models the spectrum as a directed graph. Here, spectral features serve as nodes, and a graph attention mechanism dynamically aggregates information from ``jaspectral neighbors". This allows the model to learn intra-spectral dependencies (e.g., how the intensity of a ``white line" peak correlates with the curvature of the post-edge region) through content-aware aggregation rather than fixed-weight convolution. By unifying these two representations, global scattering physics via frequency regularization and local electronic structure via graph attention, the SPT creates a robust mapping from $\mu(E)$ to $\rho(r)$ that is resilient to the noise inherent in experimental data.

\subsection{Overcoming the Instability of Spectral Inversion}

    \par To demonstrate that the SPT framework successfully resolves the ill-posed nature of XANES inversion, we conducted a rigorous benchmarking analysis against both conventional machine learning algorithms (Random Forests, Extra Trees, XGBoost) and state-of-the-art transformer models (FTTransformer, AMFormer) on a held-out test set. The results indicate that the SPT establishes a new state-of-the-art for structural prediction, consistently outperforming baseline models across diverse crystallographic environments not only in accuracy but in architectural efficiency.

    \par A critical distinction of the SPT framework is its capacity for simultaneous multi-descriptor prediction. Unlike baseline ensemble methods such as Random Forests or Gradient Boosting, which typically require the training of fragmented, disparate models for each individual structural property (e.g., separate regressors for coordination number and oxidation state), the SPT architecture constructs a unified, physics-informed latent representation. This shared feature space, enriched by the frequency-domain regularization and graph-based contextual attention, captures the holistic scattering physics of the absorbing site. Consequently, the SPT can simultaneously disentangle and predict the full suite of electronic (OS) and geometric (CN, CN2, NNRS, PSGO, see the detailed definitions in the Sec.~\ref{localstructure} of the Method part) descriptors from a single spectral input. This architectural unification prevents the physical inconsistencies often observed in single-task baseline models and enables the exploitation of synergistic correlations between descriptors, where learning the electronic state, for instance, informs and refines the prediction of geometric symmetry.

    \par As evidenced in Figure~\ref{comparion}a and b, the SPT exhibits exceptional precision in identifying metallic coordination environments. For the classification of OS, SPT achieves an F1-score of 0.969. Notably, in predicting the second coordination number (CN2), a critical parameter for defining local polyhedral geometry that is notoriously difficult to resolve, the model attained an F1-score of 0.811, significantly outstripping the Random Forests baseline (0.795). In continuous regression tasks, which require mapping subtle spectral curvature to radial distances, SPT achieved a coefficient of determination ($R^{2}$) of 0.764 for the nearest-neighbor radial standard deviation (NNRS), effectively mitigating the discretization errors inherent to tree-based ensembles.

    \par Crucially, the model's performance extends to the distinct scattering landscape of oxygen sites (Figure~\ref{comparion}d and e). SPT demonstrates remarkably high fidelity in quantifying electronic modulation, achieving an $R^{2}$ of 0.951 for the nearest-neighbor oxidation state mean (NNOSM), a substantial improvement over the baselines. This indicates that the SPT is highly sensitive to the spectral shifts induced by the oxidation states of neighboring atoms. Furthermore, it maintains high classification accuracy for local coordination (CN F1-score: 0.951), proving its capability to disentangle structural signals across chemically distinct elements.

    \par These results collectively indicate that the dual frequency-graph architecture captures physical scattering dependencies more effectively than pure self-attention mechanisms or fragmented ensemble methods. By treating the spectrum as a unified physical entity rather than a collection of independent features, the SPT distinguishes between genuine structural signals and high-frequency spectral artifacts.

    \par A solution to an ill-posed inverse problem must be stable against perturbations. In the Fredholm formulation (Eq.~\ref{eq:Fredholm}), trivial experimental noise ($\epsilon$) can amplify into catastrophic errors in the reconstructed structure $\rho(r)$. To probe this stability, we subjected the SPT to a stress test by injecting random Gaussian noise with intensities ranging from $\sigma=0.1$ to $\sigma=0.3$ into the input spectra (Figure~\ref{comparion}c and f); the details can be found in the Sec.~\ref{RobustnessAnalysis} of the Methods.

    \par While standard deep learning models and single-task baselines typically suffer rapid performance collapse under such stochastic regimes, the SPT exhibits remarkable resilience. For non-oxygen sites, the prediction of oxidation states (OS) remains robust, maintaining an F1-score of 0.910 even at the maximum noise level ($\sigma=0.30$). This stability is even more pronounced in the oxygen site dataset, where the model retains an F1-score of 0.907 for coordination numbers (CN) and an $R^{2}$ of 0.902 for oxidation state modulation (NNOSM) under severe distortion. These bounded degradations corroborate our theoretical hypothesis: the initial frequency-domain decomposition functions as an effective low-pass filter, segregating the robust, low-frequency signal components that encode fundamental chemical states from the high-frequency components where noise resides. Crucially, this stability serves as empirical evidence that the model has approximated a Lipschitz-continuous inverse operator regarding the structural descriptors, successfully disentangling the high-frequency experimental noise manifold from the physical signal manifold. Furthermore, the multi-target nature of the SPT acts as a regularizer, enforcing physical consistency across predicted descriptors and preventing the model from overfitting to noise artifacts that might mislead isolated single-task models.

    \par Although descriptors sensitive to fine long-range symmetry (e.g., PSGO) showed expected sensitivity, the overall stability of local chemical descriptors proves that the SPT has learned invariant physical laws governing the spectrum-structure relationship, rather than merely memorizing training data. A detailed comparative analysis against baseline models is provided in Supplementary Table 5 and 6.

\subsection{Generalization to Dynamic, Disordered and Experimental Regimes}

\subsubsection{Resolving Dynamic Phase Evolution in Crystalline Systems}

    \par
\begin{figure*}
        \begin{center}
            \centering
            \includegraphics[width=0.8\linewidth]{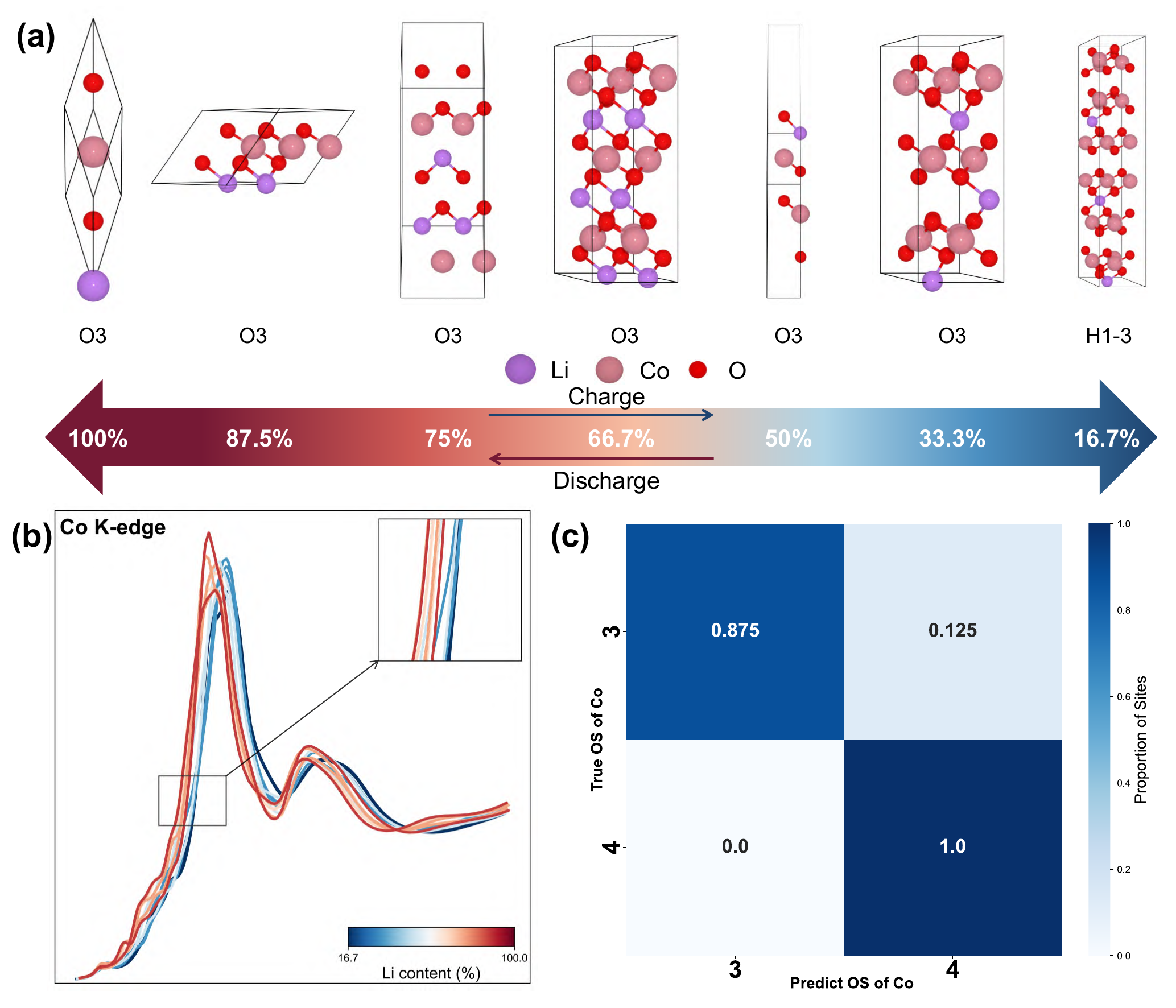}
            \caption{\textbf{SPT-enabled tracking of structural and electronic evolution in LiCoO$_2$ during delithiation}. (a) Schematic illustration of the phase transition pathway in LiCoO$_2$ as lithium content decreases, showing the transformation from the O3 phase to the H1-3 phase. (b) Co K-edge XANES spectra and corresponding oxidation states for various lithium concentrations. (c) The accuracy of SPT in distinguishing between Co$^{3+}$ and Co$^{4+}$ states across different structural phases.
            }
            \label{LiCO}
        \end{center}
    \end{figure*}

 In this section, we first apply the SPT to the Li-Co-O system to evaluate its capability in capturing the evolution of local structures and electronic states during lithium deintercalation. As depicted in Figure~\ref{LiCO}(a), the LiCoO$_{2}$ crystal undergoes a series of phase transitions with decreasing lithium content from full lithiation to deep delithiation, all structures are adopted from the previous work~\cite{Ran2021LCO}. The structure evolves from the well-ordered O3 phase at high lithium concentrations to the more complex H1-3 phase at low lithium content, reflecting significant rearrangements in the local atomic environment~\cite{qian2018electrochemical,ohnishi2021situ,yao2024scalable}.
    
    \par To quantitatively probe the electronic structure changes accompanying these phase transitions, the SPT method is employed to predict the oxidation state of cobalt ions across the entire delithiation pathway. Figure~\ref{LiCO}(b) presents the predicted valence states based on the Co K-edge XANES spectra for a range of lithium contents. The SPT accurately captures the progressive increase in cobalt oxidation state from Co$^{3+}$ to Co$^{4+}$ as lithium is extracted, in excellent agreement with the expected redox behavior. The predicted oxidation states exhibit a clear and systematic shift, demonstrating the model’s sensitivity to subtle spectral variations associated with changes in local electronic structure. The classification performance is rigorously quantified by the confusion matrix in Figure~\ref{LiCO}(c). Strikingly, SPT achieves near-perfect classification accuracy, specifically attaining a 100\% recognition rate for the Co$^{4+}$ state. This result highlights the model's robust capability to disentangle electronic signatures from significant structural distortions. Notably, the SPT maintains strong predictive performance across both the O3 and H1-3 phases, indicating its exceptional transferability and generalization to distinct crystallographic environments within the same material system.
    
    \par The application of the SPT to the Li-Co-O system demonstrates its capacity to resolve intricate electronic and structural transformations during complex phase evolution. By enabling accurate, phase-resolved prediction of oxidation states and local coordination environments, SPT provides a robust framework for elucidating dynamic charge-compensation mechanisms in functional materials. This capability is critical for advancing the real-time characterization and rational design of high-performance energy storage materials, particularly for understanding degradation pathways and stabilizing metastable phases.

\subsubsection{Generalization to Disordered and Amorphous Environments}
    
    \begin{figure*}[t]
    \begin{center}
        \centering
        \includegraphics[width=\linewidth]{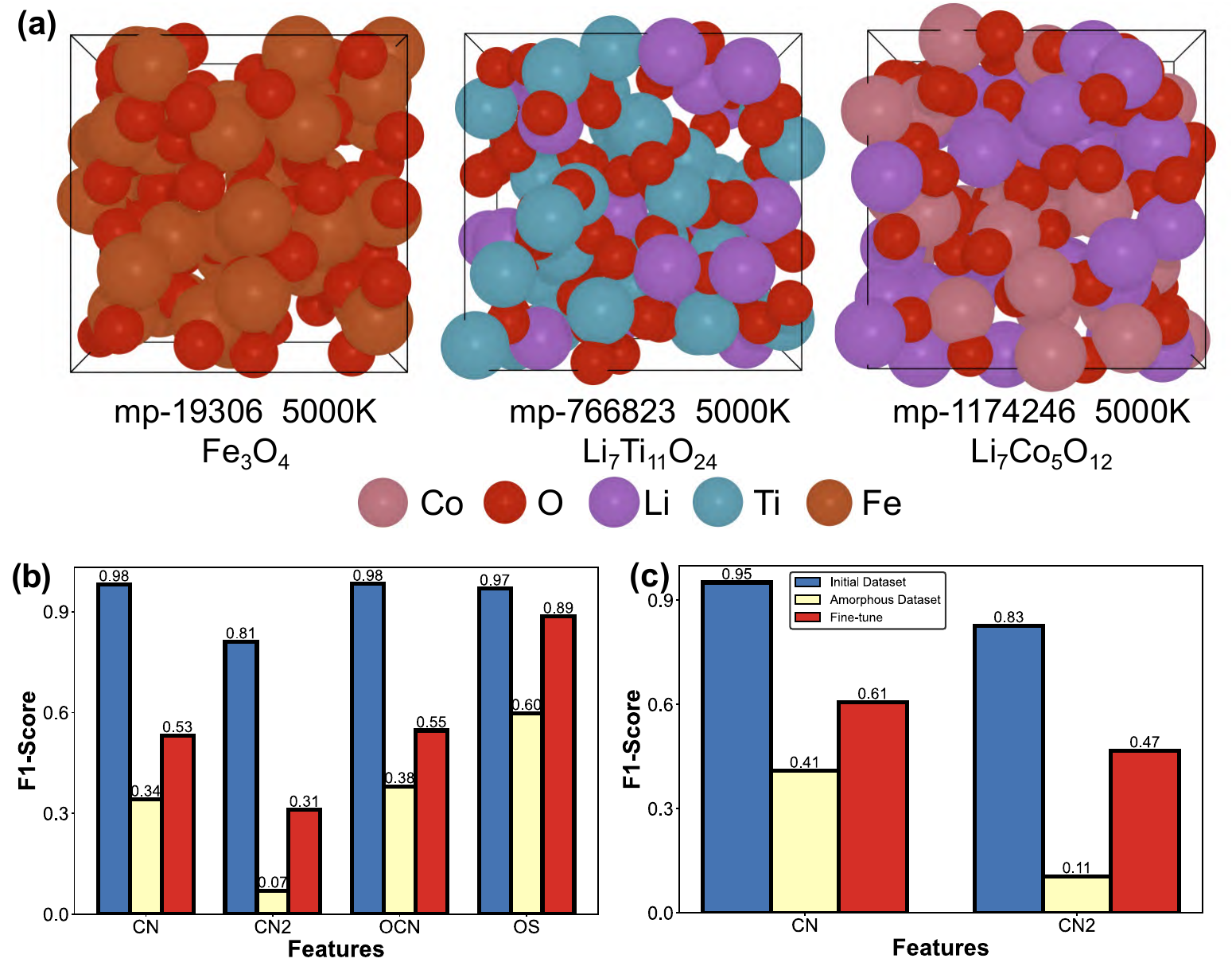}
        \caption{
            \textbf{SPT performance in transfer learning for descriptor prediction in amorphous structures.}
            (a) Atomic models of representative amorphous materials, including Fe$_3$O$_4$, Li$_7$Ti$_{11}$O$_{24}$, and Li$_7$Co$_5$O$_{12}$, generated at 5000~K. Different elements are distinguished by color, highlighting the disordered atomic arrangement and compositional diversity characteristic of amorphous phases. (b, c) Comparison of the SPT performence for structural descriptors of CN, CN2, OCN and OS across the initial crystalline-trained model (blue), direct application to the amorphous dataset (yellow), and after fine-tuning on amorphous data (red). 
        }
        \label{AM}
    \end{center}
    \end{figure*}

    \par To further assess the transferability of the SPT, we evaluate its performance in predicting local structural descriptors for amorphous materials; all structural data utilized in this evaluation are sourced from previous work~\cite{Zheng2024amorphousdata}. Amorphous metal oxides (AMOs), which lack crystalline long-range order, exhibit distinctive structural characteristics such as unsaturated coordination, short-range order, and surface dangling bonds~\cite{chen2015fractal, hong2020ultralow, tian2023disorder, liu2024selenium}. These traits confer flexible electronic properties~\cite{nai2015tailoring}, which contribute to their exceptional performance in energy and catalysis applications~\cite{kang2023recent}. The lack of long-range atomic order in amorphous structures presents a significant challenge to traditional prediction methods, which are typically optimized for crystalline systems~\cite{oganov2006crystal, liang2020cryspnet, jiao2023crystal}. This scenario exemplifies a classic domain adaptation problem, where the prediction task remains unchanged while the input domain—specifically the configurational space of atomic arrangements—differs significantly between the crystalline source data and the amorphous target data.
    
    \par As shown in Figure~\ref{AM}(a), we generate atomic models for several representative amorphous materials at 5000 K, including Fe$_3$O$_4$, Li$_7$Ti$_{11}$O$_{24}$, and Li$_7$Co$_5$O$_{12}$. With atomic species color-coded for visual distinction, these models highlight the disordered atomic arrangements and compositional diversity typical of amorphous phases. To bridge this distributional shift within the configurational domain, we employ a transfer learning strategy. By using the structural representations learned from the crystalline source domain, this paradigm allows the model to effectively adapt to the disordered target domain. This approach addresses the core challenge of domain adaptation by aligning the feature distributions of the source and target data, thereby enhancing task-specific performance even when amorphous training data is scarce. Transfer learning has been widely used in fields such as computer vision and natural language processing, demonstrating exceptional efficacy in similar scenarios~\cite{chen2023vlp, zhang2024vision, xiao2024principled}.

    \par Initially, we train an SPT on the crystalline dataset from the source domain to learn generic representations of X-ray absorption spectra. This model is then directly applied to the target domain amorphous structures to assess its zero-shot performance. Subsequently, we fine-tuned the pre-trained model using a small amorphous dataset. This approach allows the model to retain robust features learned from the large crystalline dataset while adjusting parameters to capture the local atomic environments and structural details of the amorphous phases. Figure~\ref{AM}(b) and (c) present a comparative analysis of F1-Scores specifically for non-oxygen sites and oxygen sites, respectively. When the initial crystal-trained model is directly applied to amorphous datasets, predictive performance drops significantly due to the domain gap. However, after fine-tuning, the SPT recovers considerable predictive accuracy. Notably, the prediction of OS for non-oxygen sites exhibits a robust recovery, rebounding from 0.60 to 0.89. For geometric descriptors such as CN, CN2, and OCN, the fine-tuning process also yields distinct improvements; for instance, the F1-Score for CN in non-oxygen sites increases from 0.34 to 0.53. Nevertheless, compared to the OS, the recovery for these geometric features remains less optimal. This disparity suggests that mapping spectral features to the complex, disordered topologies of amorphous structures requires a significantly denser sampling of the chemical space. We anticipate that scaling the amorphous training corpus will markedly improve the performance of these geometric descriptors, as the model learns to disentangle subtle structural variations from larger datasets. Nevertheless, it is also critical to consider the intrinsic limitations of the descriptors themselves: certain rigid crystallographic definitions may be ill-suited for the continuous random networks of amorphous systems, suggesting that future work should not only expand data but also re-evaluate the applicability of discrete descriptors in disordered regimes.

    \par As we show, the fine-tuned SPT exhibits strong adaptability to the unique atomic arrangements and compositional heterogeneity of amorphous materials. Crucially, the SPT aims to resolve explicit physical descriptors of the local environment, rather than attempting a reconstruction of full-atomic coordinates. This distinction highlights the physical basis of the model's generalization: rather than simply memorizing the data manifold of the crystalline training set, the SPT, facilitated by its frequency-domain regularization, approximates the inverse mapping of the local scattering operator. By capturing this intrinsic invariance of scattering physics, the model yields reliable descriptor predictions even across the domain shift to disordered systems. This versatility highlights the potential of the SPT for characterizing and designing complex amorphous materials.



\subsubsection{Bridging the Simulation-Experiment Gap}

    \begin{figure*}[t] 
    \centering
    \includegraphics[width=\linewidth]{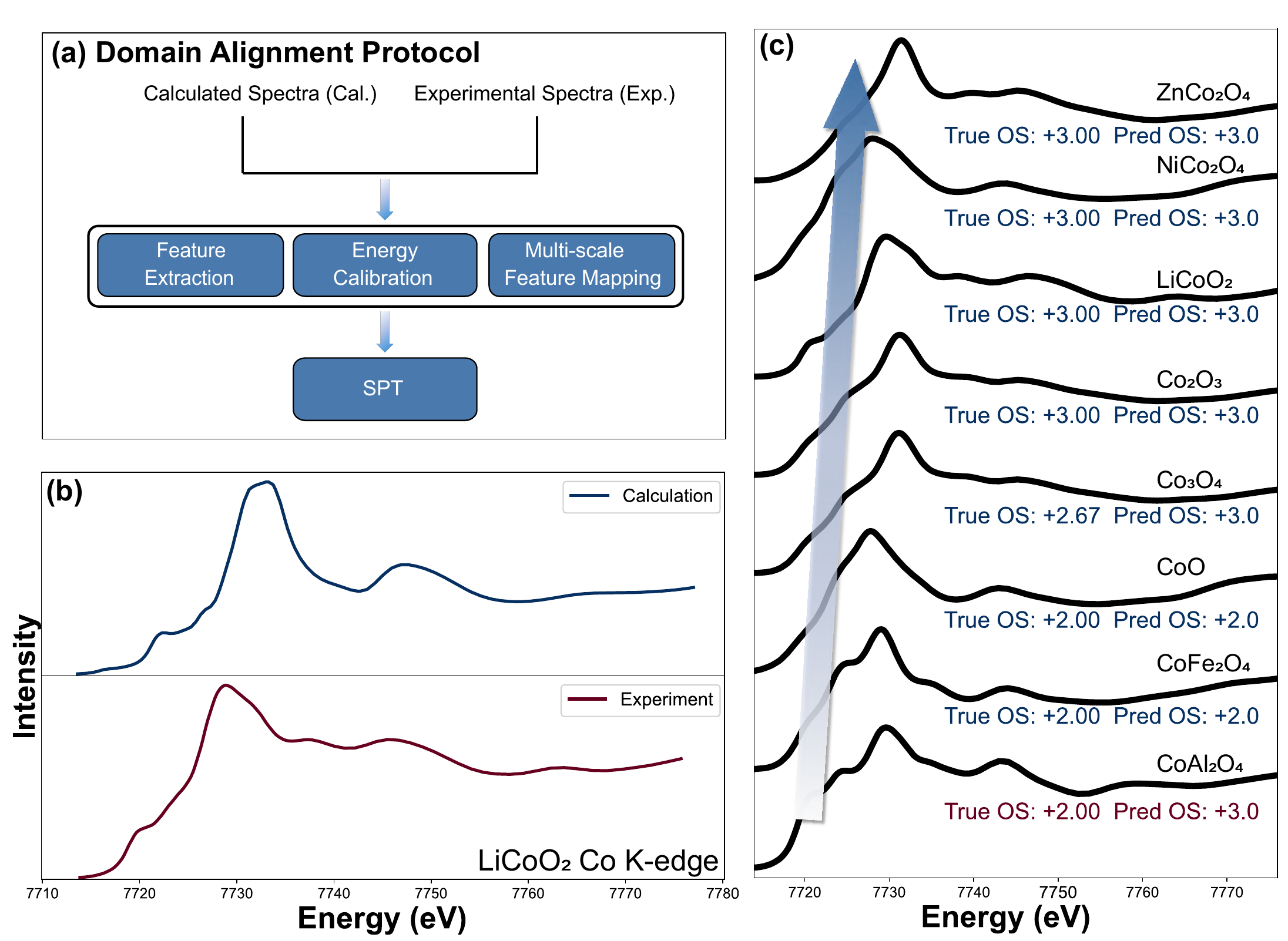} 
    \caption{
        \textbf{Generalization assessment of SPT framework on experimental cobalt K-edge XANES spectra.}
        (a) Schematic of the domain alignment protocol. The workflow illustrates the bridging of theoretical calculations (Cal.) and experimental measurements (Exp.) via multi-scale feature extraction and mapping, ensuring distributional consistency between domains.
        (b) Spectral energy calibration. A representative comparison between the calculated and experimental K-edge spectra of $\mathrm{LiCoO_2}$, demonstrating the precise alignment of absorption edges (approx. 7720 eV) required for valid inference.
        (c) Prediction performance on diverse cobalt oxides. The subpanels annotate the ground truth versus predicted oxidation states.
    }
    \label{fig:exp_xas}
    \end{figure*}

    \par The ultimate validation of the SPT framework lies in its ability to bridge the ``simulation-reality gap" and generalize to experimental data. Although trained exclusively on theoretical datasets, the SPT model demonstrates a remarkable capacity to track structural evolution within real-world chemical processes, driven by the universality of the underlying physical laws governing structure-spectrum relationships. To facilitate this transfer, we implemented a rigorous domain alignment protocol, shown in Figure~\ref{fig:exp_xas}(a) and (b). By incorporating multi-scale feature extraction and precise energy calibration, this protocol effectively mitigates the impact of experimental noise and instrumental broadening, ensuring that input spectral features remain highly faithful to the learned training distribution.

    \par As illustrated in Figure~\ref{fig:exp_xas}(c), the framework exhibits exceptional generalization capabilities across a diverse library of cobalt oxides, successfully deciphering oxidation states in systems with varying coordination environments: The model achieved high-precision classification for standard octahedral geometries, accurately identifying the +3 oxidation state in layered LiCoO$_{2}$, Co$_{2}$O$_{3}$, and the spinel ZnCo$_{2}$O$_{4}$; While in the divalent systems, it demonstrated equal robustness in identifying the +2 oxidation state in rock-salt CoO and the inverse spinel CoFe$_{2}$O$_{4}$. Meanwhile, the SPT model shows a high sensitivity to complex topological features: For example, in systems such as CoAl$_{2}$O$_{4}$, the model output showed slighty deviations from nominal integer valence states: In the case of CoAl$_{2}$O$_{4}$, the cobalt occupies tetrahedral sites, a geometry distinct from the dominant octahedral environments in the training set. The model's behavior in analyzing complex cation distributions further underscores its sensitivity to subtle topological variations rather than mere limitations. Rather than indicating a failure of the learning algorithm, these deviations highlight the acute sensitivity of the SPT's to specific coordination geometries. The model correctly detects that these spectra do not conform to standard octahedral fingerprints. This suggests that the framework's predictive scope can be readily expanded to include these complex topologies by incorporating a broader diversity of coordination geometries into the training phase.

    \par In short, the SPT framework proves to be a powerful tool for experimental analysis. It successfully neutralizes the interference of experimental noise to retrieve accurate chemical information for the vast majority of standard material systems, establishing a solid foundation for high-throughput, automated interpretation of X-ray absorption spectroscopy.

\section{Discussion}

\par The ``solution'' to an inverse problem in physics requires more than state-of-the-art prediction accuracy; it demands the capture of underlying physical invariance. The SPT framework validates this through its zero-shot generalization to amorphous systems. Since local scattering physics is invariant across crystalline and amorphous phases, the model's ability to transfer knowledge across these topological distinct domains proves it has learned the intrinsic inverse scattering rules rather than memorizing specific crystalline patterns. By redefining the target as robust statistical descriptors and enforcing physical consistency via frequency-domain regularization, SPT transforms the originally ill-posed XAS inversion into a solvable, deterministic computational process.

\par This work establishes a transformative, data-driven paradigm for XAS, effectively bridging the decades-old ``simulation-reality gap" between first-principles theory and experimental measurement. By synergizing physics-informed frequency-domain decomposition with graph-based contextual attention, the SPT transcends the limitations of traditional ``black-box" models. It hierarchically captures both global spectral morphology and localized electronic fluctuations, providing a robust solution to the notoriously ill-posed inverse problem of XANES. The cornerstone of this framework is the integration of physical laws as an inductive bias, specifically the Fourier duality between energy-space oscillations and radial atomic scattering. This approach allows the model to ``diagonalize" complex scattering kernels, segregating robust structural signals from destabilizing experimental noise. Supported by our foundational XAS Data Engine containing over 50,000 unique atomic sites, the SPT achieves state-of-the-art accuracy, consistently outperforming both ensemble learning and pure transformer-based architectures. More importantly, its remarkable resilience to noise and its zero-shot generalization, recovering performance in amorphous systems with minimal fine-tuning, demonstrate that the model has captured invariant physical correlations rather than merely memorizing local training data. As such, the SPT represents a robust, interpretable, and unified engine for spectroscopic inference.

\par Beyond its immediate utility as an interpretative solver, the SPT framework represents a critical architectural advance toward autonomous materials discovery~\cite{Burger2020roboticchemist, Zhang2025AI, Song2025AIChemist}. Traditional spectroscopic analysis reliant on human expertise or computationally expensive DFT has long been the rate limiting step in high-throughput experimentation. In contrast, the SPT’s ability to infer structural descriptors with millisecond-scale latency effectively removes this bottleneck, allowing it to function as the ``eyes" of closed-loop robotic chemists. By integrating SPT into automated synthesis pipelines, self-driving laboratories can receive real-time structural feedback on synthesized intermediates, enabling on-the-fly trajectory optimization without human intervention. As the ``AI for Science" paradigm evolves toward universal architectures, the SPT provides a strategic blueprint for a Spectroscopic Foundation Model. The learned frequency-domain representations capture universal scattering physics that can potentially extend to other core-level techniques, such as EELS or XPS. Ultimately, this framework serves as a vital component of the Materials Project Initiative, accelerating the discovery of next-generation functional materials through the seamless integration of high-throughput theoretical insights and real-world experimental reality.

 \section{Methods}
 
    \subsection{Dataset Generation}
    \par We select multi-component transition metal oxides (TMOs), like Nickel-Manganese-Cobalt oxides (NMCs) and lithium iron phosphate (LFP) type materials, as our sample systems due to their prevalence in energy storage technologies, particularly lithium-ion batteries~\cite{wu2024ni,ji2023direct,jang2024structurally,marie2024trapped}. In these materials, electrochemical properties such as redox activity, capacity, and stability are governed by the interactions between constituent elements and their local atomic environments.

    \par Crystal structures are obtained from the Materials Project~\cite{jain10materials}, encompassing a comprehensive set of systems within the Co-Li-Mn-Ni-O chemical space and related subsets. These structures contain over over 3000 crystal structures and 50,000 nonequivalent sites. The choice of the computational engine was governed by two opposing constraints: 1) Physical accuracy: The simulation must capture the essential physics of the photoabsorption process, including multiple scattering events and electronic transitions to unoccupied states; 2) Computational throughput: The method must be efficient enough to compute tens of thousands of spectra within a reasonable timeframe. We utilize the FEFF10 code~\cite{Rehr2021}, which employs a real-space multiple scattering formalism based on the Green's function method. This approach calculates the X-ray absorption coefficient $\mu(E)$ by summing over all scattering paths of the photoelectron within a cluster of atoms centered on the absorber with a reasonable agreement with experimental data~\cite{mathew2018high,Rehr2021}. Further details on the computational parameters are provided in the Supplementary Information (SI).

    \par
\begin{figure*}
		\begin{center}
		\centering
		\includegraphics[width=\linewidth]{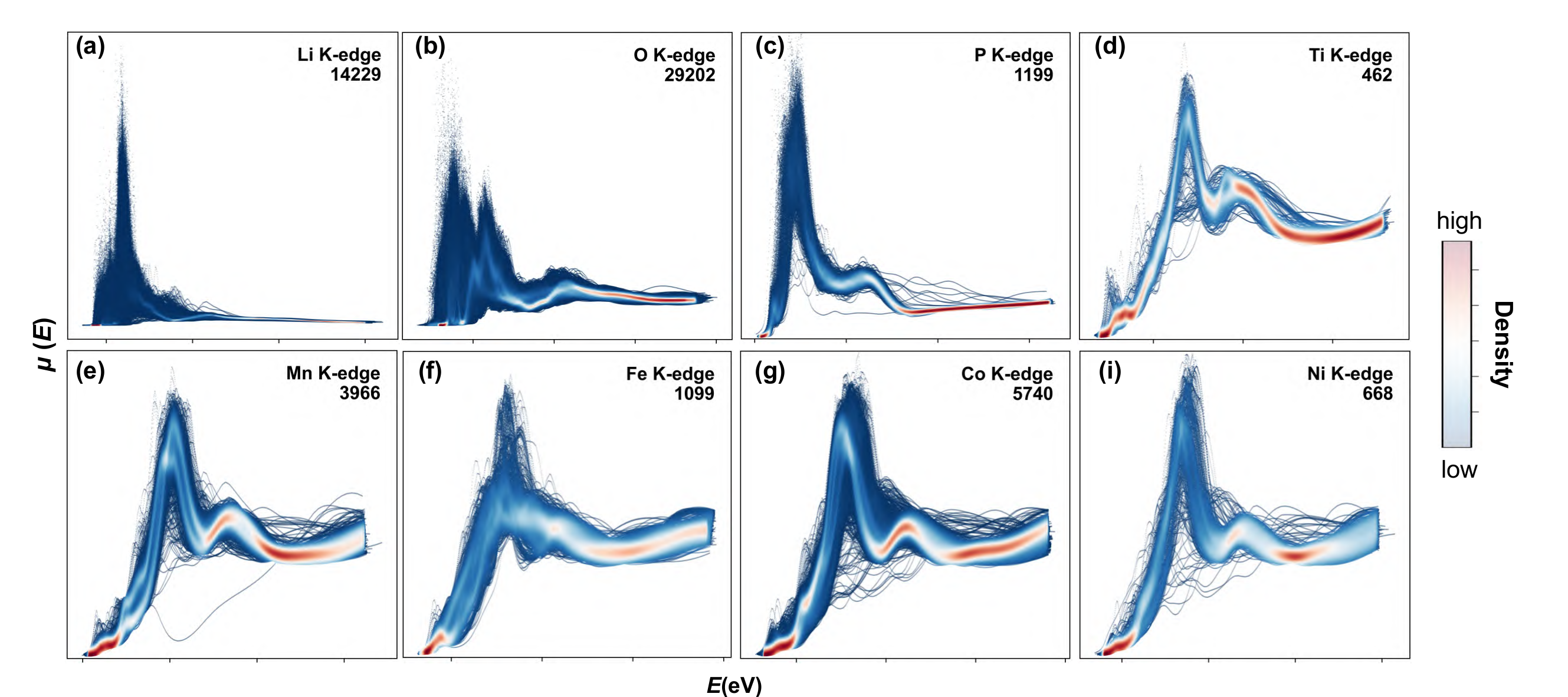}
		\caption{\textbf{Distribution of computed K-edge XANES spectra across eight elemental species.}
        Each panel displays the spectral density for a specific element (Li, O, P, Ti, Mn, Fe, Co, Ni) within the dataset, with the total count of constituent spectra annotated. The horizontal axis represents relative energy (eV) with respect to the absorption edge, while the vertical axis denotes normalized absorption intensity. The color gradient indicates the density of spectral lines, highlighting the dominant spectral features for each element.}
		\label{XAS_Dis}
		\end{center}
    \end{figure*}

 To ascertain the breadth of the data distribution within the dataset, we analyze the distribution of X-ray absorption spectra across the different elements. Figure~\ref{XAS_Dis} presents these distributions, with each panel corresponding to a specific element's K-edge. The horizontal axis denotes the relative energy, while the vertical axis represents the absorption intensity. The overall spectral shapes are similar among the transition metals, and the pre-edge peak intensity diminishes for the later 3$d$ transition metals, consistent with a previous report~\cite{PhysRevMaterials.3.033604}. In contrast, the Lithium K-edge XANES exhibits dense pre-edge and main-edge features. The Phosphorus K-edge spectra show little variation, which can be attributed to the similar chemical environment of phosphorus atoms within the LFP materials structures. Conversely, the Oxygen K-edge spectra contain rich information, reflecting the diverse chemical environments of oxygen across the various oxides.

    \subsection{Local Structure Descriptors}\label{localstructure}

    \par Understanding the local atomic environments in TMOs is crucial for connecting atomic-scale structure to electrochemical performance. For experimentalists, tracking changes in the local coordination of transition metal, lithium, and oxygen atoms provides critical insights into how structural variations govern redox behavior, capacity retention, and material stability~\cite{yang2021situ,liu2024tailoring}. Here, we characterize these environments using a set of structural descriptors derived from chemical intuition. Specifically, we compute six features for oxygen sites and seven for non-oxygen sites, capturing key structural and electronic information within the second coordination shell of the absorbing atom, consistent with established methodologies~\cite{liang2023decoding}.
    
    \begin{figure}
		\begin{center}
		\centering
		\includegraphics[width=\linewidth]{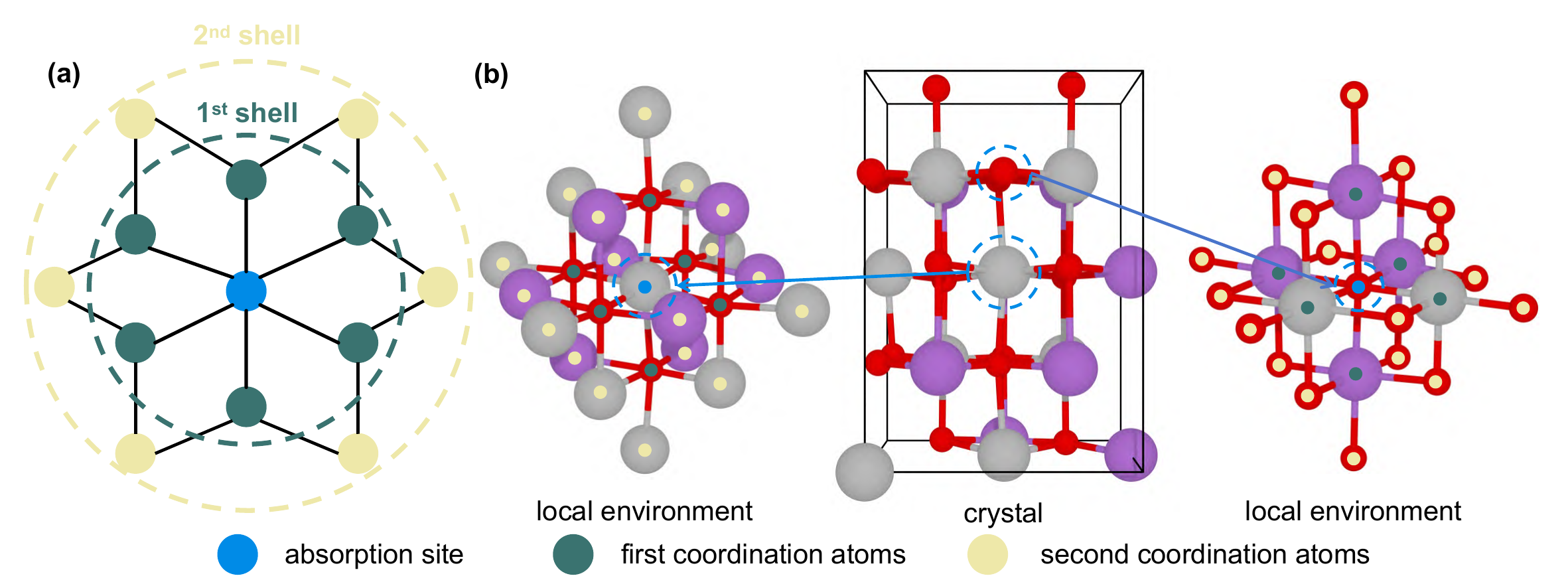}
		\caption{\textbf{Schematic of the local atomic environment.}
        (a) Schematic illustration of the absorbing site and local environment. (b) Local atomic structures of oxygen and non - oxygen sites, with absorbing atoms marked in blue, first nearest-neighbor atoms in green, and second nearest-neighbor atoms in light yellow.}
		\label{local_env}
		\end{center}
    \end{figure}

    \paragraph{For oxygen sites:}
    \begin{itemize}
        \item \textbf{OS:} Integer oxidation states of the absorbing site are determined using the BVAnalyzer from the PYMATGEN package~\cite{ong2013python}.
        \item \textbf{CN:} Number of first nearest-neighbor coordinating atoms at the absorbing site, determined using the CrystalNN from the PYMATGEN package~\cite{ong2013python}.
        \item \textbf{CN2:} Number of nearest-neighbor atoms to the absorbing site's first nearest-neighbor coordinating atoms, as illustrated in Figure~\ref{local_env}(a), corresponding to the atoms marked in light yellow in Figure~\ref{local_env}(b).
        \item \textbf{NNRS:} Standard deviation of bond lengths between the absorbing site atom and its nearest-neighbor atoms.
        \item \textbf{NNOSM:} Average oxidation state of the nearest-neighbor atoms.
        \item \textbf{PSGO:} Point group symmetry order represented by an integer, computed as the total number of symmetry operations for the point group of the full crystal space group, determined using the PYMATGEN package~\cite{ong2013python}.
    \end{itemize}
    
    \paragraph{For non-oxygen sites:}
    The definitions of \textbf{OS}, \textbf{CN}, \textbf{CN2}, \textbf{NNRS}, and \textbf{PSGO} are consistent with those for the oxygen sites.
    \begin{itemize}
        \item \textbf{OCN:} The number of oxygen atoms among the first nearest-neighbor coordinating atoms.
        \item \textbf{MOOD:} The minimum oxygen–oxygen distance of the first nearest oxygen atoms.
    \end{itemize}

   \subsection{SPT Model Architecture}

   \par The XAS spectral data contains substantial redundant information, which creates significant interference in establishing effective mappings between spectral signatures and target descriptors. Conventional analytical methodologies face two fundamental limitations in processing XAS spectral data. Firstly, holistic analysis approaches emphasizing spectral averaging tend to obscure subtle local features near absorption edges. Moreover, segmented analysis techniques focusing on specific spectral regions inadequately characterize long-range correlations spanning the complete spectral range~\cite{yang2021situ, guda2021understanding, cutsail2022challenges}. 
    
    \par We propose a Spectral Pattern Translator (SPT) method for predicting the structural descriptors of XAS spectra, which is capable of simultaneously focusing on the local detailed fluctuations and the overall trend changes of the spectra, and realizes the hierarchical extraction of complex spectral features. Inspired by modelling approaches in the field of time series analysis~\cite{wu2022timesnet, liu2023itransformer, ekambaram2023tsmixer}, the SPT decomposes the spectral information into multiple frequency bands and guides the fusion of tensors with different frequency characteristics through a set of spectral queries projected from the semantic space. For the spectral local structure, we construct a mechanism for modeling the association of nodes between neighboring regions. 
    
    \subsubsection{Multi-scale Spectral Semantic Discovery}
    
    As shown in Figure~\ref{fig:model}, to mitigate interference from high-frequency noise and isolate physically significant scattering modes, we decompose the spectral tensors into multi-frequency components via spectral decomposition and encode them into 2D tensor groups. Subsequently, a 2D spectral convolution is applied to each frequency band to achieve hierarchical feature extraction. To establish an effective mapping between spectral semantics and target structural descriptors, we project learnable spectral query vectors that dynamically regulate the feature fusion process through frequency-domain attention weighting. This mechanism automatically amplifies features strongly correlated with the target descriptors while suppressing artifacts from redundant spectral components.
    
    Formally, let $\mathbf{A} \in \mathbb{R}^{F \times D}$ denote the input spectral embedding. We employ via the FFT to identify $K$ significant frequency bands $\Omega = \{\omega_1, \ldots, \omega_K\}$, where each band corresponds to a distinct scattering path length. First, to facilitate hierarchical feature extraction, we employ a tensorization operator $\mathcal{R}(\cdot)$ to restructure the spectral representation into a 3D tensor $\mathcal{T} \in \mathbb{R}^{K \times P \times D}$ with period length $P$. The spectral features $\mathbf{Z}_k$ for the $k$-th band are then extracted using a 2D convolution kernel $\mathbf{W}_k$. Meanwhile, a learnable query vector $\mathbf{q} \in \mathbb{R}^{K}$ is projected from the global context to compute frequency-aware attention scores $\boldsymbol{\alpha}$ via a softmax layer. Finally, the refined spectral sequence $\mathbf{A}'$ is reconstructed through the inverse reshaping operator $\mathcal{R}^{-1}$ after weighted aggregation, as Eq.~\eqref{eq:spectral_conv}, Eq.~\eqref{eq:attention_weights}, Eq.~\eqref{eq:aggregation}:
    
    \begin{align}
        \label{eq:spectral_conv}
        \mathbf{Z}_k &= \mathbf{W}_k \ast \mathcal{R}_k \left( \mathcal{F}(\mathbf{A}) \right), \quad \forall k \in \{1, \ldots, K\} \\
        \label{eq:attention_weights}
        \boldsymbol{\alpha} &= \sigma \left( \phi(\mathbf{A})^\top \mathbf{q} \right) \\
        \label{eq:aggregation}
        \mathbf{A}' &= \mathcal{R}^{-1} \left( \sum_{k=1}^{K} \alpha_k \cdot \mathbf{Z}_k \right)
    \end{align}
    
    \noindent where $\mathcal{F}(\cdot)$ denotes the FFT operator. In Equation~\ref{eq:attention_weights}, $\phi(\cdot)$ represents a global context mapping parameterized by a Multi-Layer Perceptron, and $\sigma(\cdot)$ is the softmax function.

    \subsubsection{Aggregate Spectral Neighbourhood Information}
    
    While the frequency-domain analysis captures global scattering modes, spectral perturbations induced by local electronic events—such as charge transfer or Jahn-Teller distortions—manifest as correlated shifts across specific energy ranges. To capture these local dependencies, we model the refined spectral sequence as a directed graph $\mathcal{G}=(\mathcal{V}, \mathcal{E})$. We define the node feature matrix $\mathbf{H} = \mathbf{A}' \in \mathbb{R}^{T \times D}$, where each node $h_i$ corresponds to a spectral point at energy $E_i$.

    To enforce physical locality, we construct the edge set $\mathcal{E}$ based on a sliding window mechanism. An edge directed from $j$ to $i$ exists if and only if the spectral distance satisfies $|i-j| \le M$, defining a local neighborhood set $\mathcal{N}_i$. We then introduce a graph attention mechanism to dynamically aggregate contextual information. For every connected pair $(i, j)$, we compute a pair-wise relevance score $e_{ij}$ that quantifies the influence of the spectral neighbour $j$ on the target $i$:
    
    \begin{equation}
        \label{eq:graph_attention_score}
        e_{ij} = \mathbf{v}^\top \tanh \left( \mathbf{W}_e [ \mathbf{h}_i \, \| \, \mathbf{h}_j ] \right)
    \end{equation}
    
    \noindent where $\|$ denotes the concatenation operation, $\mathbf{W}_e \in \mathbb{R}^{D' \times 2D}$ is a learnable linear transformation that projects the concatenated features into a latent interaction space, and $\mathbf{v}$ is a weight vector.
    
    To normalize these interactions across the local neighborhood, we apply a softmax function constrained by the graph topology. The updated representation $\tilde{\mathbf{h}}_i$ for each spectral node is computed as a weighted sum of its neighbours:
    
    \begin{align}
        \label{eq:graph_aggregation}
        \alpha_{ij} &= \frac{\exp(e_{ij})}{\sum_{k \in \mathcal{N}_i} \exp(e_{ik})} \\
        \tilde{\mathbf{h}}_i &= \sigma \left( \sum_{j \in \mathcal{N}_i} \alpha_{ij} \mathbf{W}_v \mathbf{h}_j \right)
    \end{align}
    
    \noindent where $\alpha_{ij}$ represents the normalized attention coefficient indicating the importance of feature $j$ to feature $i$, and $\mathbf{W}_v$ is a value projection matrix. By explicitly constraining the aggregation within $\mathcal{N}_i$, this mechanism effectively acts as a spectral proximity filter, adaptively smoothing the signal based on local electronic correlations while preserving sharp features essential for structural identification.

  \subsection{Robustness Analysis of SPT}\label{RobustnessAnalysis}

      \begin{figure*}[t!]
        \centering
        \includegraphics[width=\linewidth]{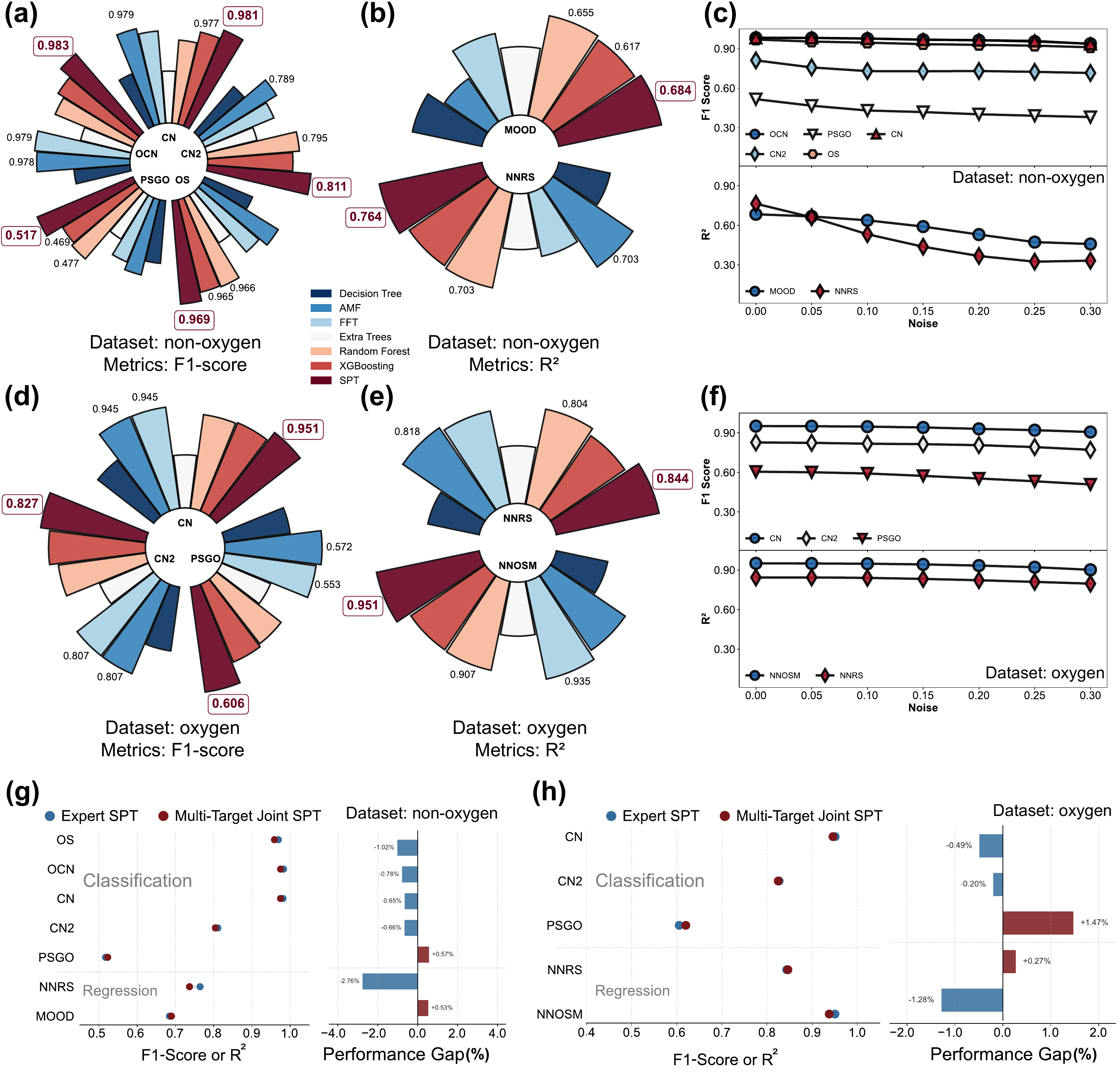}
        \caption{
            \textbf{Performance benchmarking and robustness analysis of the SPT framework.}
            (a, b) Comparative evaluation of SPT against baseline models (Decision Tree, Extra Trees, Random Forests, XGBoost, FTTransformer, and AMFormer) on non-oxygen datasets. Metrics include F1-Score for classification tasks and $R^2$ for regression tasks.
            (c) Robustness of SPT on non-oxygen sites under varying Gaussian noise intensities within the range of $[-0.3, +0.3]$.
            (d, e) Performance comparison on oxygen site datasets for classification and regression.
            (f) Robustness analysis for oxygen sites.
            (g, h) Efficiency-accuracy trade-off analysis comparing the Expert SPT (blue) and Multi-Target Joint SPT (red) strategies. Panels correspond to (g) non-oxygen and (h) oxygen datasets. 
        }
        \label{comparion}
    \end{figure*}
    
    The analysis of real-world experimental data is frequently complicated by inherent noise and background fluctuations. To determine whether the SPT framework can accurately distinguish genuine structural variations from such artifacts, we conducted a robustness analysis by superimposing random Gaussian noise within the range of $[-0.3, +0.3]$ onto the raw XANES spectra. Notably, noise concentrated around critical absorption edges or characteristic peak regions markedly distorts the signal morphology, challenging the discrimination and fitting capabilities of the model. Figure~\ref{comparion}c and Figure~\ref{comparion}f illustrate the trends in classification (F1-Score) and regression ($R^2$) performance for non-oxygen and oxygen site datasets, respectively, as the noise intensity increases.
    
    For non-oxygen site descriptors, the SPT model demonstrates exceptional stability, particularly in classification tasks, as shown in Figure~\ref{comparion}c. Detailed statistics in Supplementary Table 5 reveal that the F1-scores for OS, OCN, and CN descriptors exhibit only a marginal decline even under substantial spectral perturbations. Specifically, the OS descriptor maintains a remarkable F1-score of 0.9102, dropping from an initial 0.969, even at the maximum noise level of $\pm 0.30$. This persistence indicates that oxidation-related spectral signatures possess strong resilience to background fluctuations, effectively preserving core electronic information. Similarly, the OCN and CN descriptors sustain performance levels exceeding 0.93 at the $\pm 0.30$ noise level, suggesting that short-range coordination information remains well-preserved amidst disturbances. In the context of regression tasks, the $R^2$ values for MOOD and NNRS descriptors display a monotonically decreasing trend with increasing noise. Under the most rigorous noise conditions, random perturbations in local spectral regions cause significant distortion of spectral morphology. This distortion undermines the consistency of regression descriptors and consequently hinders the ability of the model to accurately extract quantitative structural information. Moreover, the slightly superior robustness of MOOD suggests that it captures more global average spectral patterns, whereas NNRS primarily reflects characteristics of local bond length fluctuations.
    
    This robustness is even more pronounced within the oxygen site dataset, as depicted in Figure~\ref{comparion}f and Supplementary Table 6. The CN descriptor exhibits outstanding resilience, with the F1-score remaining above 0.90 under the strongest noise conditions. In contrast, features involving medium-to-long-range structural correlations, such as CN2, exhibit a discernible decline in performance, where the F1-score decreases from 0.7576 to 0.7155. This trend reflects the fact that spectral features corresponding to outer coordination shells are inherently more subtle and thus slightly more susceptible to noise interference than the dominant first-shell signals. Most strikingly, the SPT achieves high precision for regression tasks on oxygen sites even at severe noise levels. Unlike the non-oxygen case, the NNOSM and NNRS descriptors maintain distinctively high $R^2$ values of 0.9017 and 0.7968 at $\pm 0.30$ noise, respectively. These results imply that the distinct spectral signatures of oxygen environments in transition-metal oxides provide a stable anchor, allowing the SPT to generalize effectively and capture global structural patterns against moderate-to-high spectral distortions.
    
    Overall, this analysis confirms the resilience of the SPT framework against spectral noise. The observed performance degradation is smooth and well-bounded without abrupt instability under low-to-medium noise regimes. This indicates that the model has successfully learned invariant spectral representations that capture essential physical correlations rather than overfitting to fine, noise-sensitive details. These findings underscore the potential of the model for real-world applications where experimental XAS data inevitably contain noise.

    \begin{figure*}[t!]
		\begin{center}
		\centering
		\includegraphics[width=\linewidth]{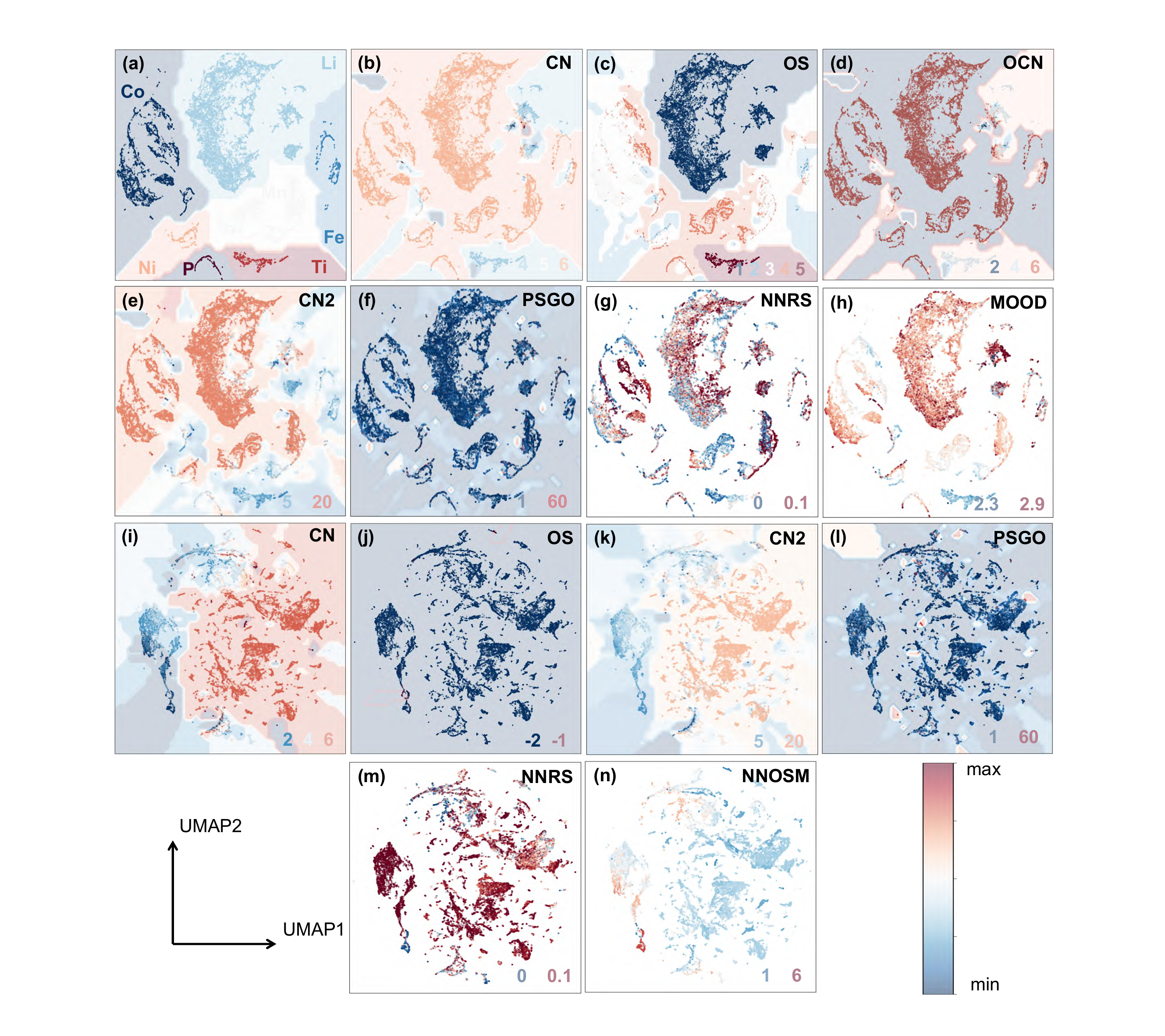}
		\caption{
            \textbf{Unsupervised visualization of spectral-structural correlations via UMAP.}
            (a) UMAP projection of the spectral dataset labeled by absorbing element, revealing distinct elemental clustering.
            (b-h) Distribution of seven local descriptors for non-oxygen sites projected onto the spectral manifold.
            (i-n) Distribution of six local descriptors for oxygen sites.
            The axes of UMAP1 and UMAP2 represent the 2D embedding of the high-dimensional spectral data. The color scale indicates the normalized value for each descriptor, demonstrating that spectra from similar local environments cluster together.
        }
        \label{umap}
		\end{center}
    \end{figure*}
    
 \subsection{Unsupervised Exploration of Spectral-Structure Correlations via UMAP}
    
    To elucidate the intricate relationship between XANES spectral features and the underlying local atomic environments, we employ the Uniform Manifold Approximation and Projection (UMAP) technique~\cite{mcinnes2018umap}. This non-linear dimensionality reduction method projects the high-dimensional spectral data into an interpretable two-dimensional manifold, defined by the UMAP1 and UMAP2 axes, as depicted in Figure~\ref{umap}. This visualization provides a powerful tool for exploring the latent structure within our dataset and for understanding how spectra cluster based on their associated physical and chemical properties.
    
    The resulting UMAP embedding reveals a highly structured organization of the spectral data, which correlates strongly with key structural and electronic descriptors. Figure~\ref{umap}(a) illustrates the distribution of various absorbing metal atoms, where elements such as Li, Ti, and Ni form distinct, well-separated clusters. This demonstrates that the inherent differences in the electronic configurations of these elements give rise to unique and distinguishable spectral fingerprints that are effectively captured by the UMAP projection.
    
    Furthermore, we systematically analyzed the distribution of local environmental descriptors across this learned manifold. Figures~\ref{umap}(b)-(h) map the seven descriptors calculated for non-oxygen sites, including CN, CN2, OS, NNRS, PSGO, OCN, and MOOD. Concurrently, Figures~\ref{umap}(i)-(n) visualize the six descriptors for oxygen sites, which include OS, CN, CN2, NNRS, NNOSM, and PSGO.
    
    A critical observation is the presence of smooth, continuous color gradients for many of the descriptors. For instance, properties like coordination number and oxidation state exhibit gradual variations across the UMAP space rather than random scattering. This continuity provides strong evidence that the UMAP embedding successfully preserves the topological structure of the original data. In other words, spectra originating from atoms in similar local environments are mapped to proximate points in the low-dimensional space, confirming a systematic and learnable correlation between spectral features and their underlying structural descriptors.

\section{Extended Discussions}
    
\subsection{Evaluation and Comparison of SPT}\label{Evaluation}
    
    \par To comprehensively evaluate model performance in predicting local structural features, we conduct a comparative analysis using several representative machine learning algorithms, including Extra Trees, Decision Tree, Random Forests, and XGBoost, alongside FTTransformer and the state-of-the-art AMFormer model ~\cite{breiman2001random,chen2016xgboost,bentsen2023spatio,cheng2024arithmetic}. All models are trained and tested on two independent datasets corresponding to non-oxygen and oxygen sites.
    
    \par Figure~\ref{comparion}(a) and (b) visually encapsulate the comparative performance across classification and regression tasks for non-oxygen sites. The SPT model exhibits a distinct performance dominance, consistently forming the outermost envelope in the radar charts across all metrics. For classification tasks involving local coordination environments (OS, CN, OCN), SPT demonstrates near-perfect fidelity, achieving F1-Scores of 0.969 for OS and 0.983 for OCN, significantly surpassing the closest competitive baselines. Even for the more challenging CN2 and PSGO descriptors, which necessitate capturing medium-to-long-range structural correlations, SPT maintains a substantial lead, achieving an F1-score of 0.811 for CN2, whereas conventional models struggle to extract these subtle spectral features. In the regression regime (Figure~\ref{comparion}(b)), the superiority of SPT is even more pronounced. It achieves an $R^2$ of 0.764 for NNRS and 0.684 for MOOD, effectively resolving continuous geometric distortions—a task where discrete decision-tree ensembles often suffer from discretization errors.

    \par The oxygen site dataset presents a more complex coordination standard, serving as a critical testbed for model generalization. As illustrated in Figure~\ref{comparion}(d) and (e), SPT retains its architectural advantage, delivering robust predictions where baseline models exhibit noticeable performance degradation. In classification tasks (Figure~\ref{comparion}(d)), SPT achieves exemplary F1-Scores of 0.951 for CN and 0.827 for CN2, underscoring its capability to disentangle first- and second-shell coordination environments with high precision. Most strikingly, in the regression tasks for oxygen sites (Figure~\ref{comparion}(e)), the model demonstrates an exceptional ability to quantify electronic modulation. Specifically, SPT attains a remarkable $R^2$ value of 0.951 for the NNOSM descriptor, outstripping the best-performing baseline (FTTransformer, $R^2=0.935$) and far exceeding ensemble methods. This result highlights that the deep feature extraction mechanism of SPT effectively captures the nuanced spectral shifts associated with neighboring atomic oxidation states, providing a highly reliable tool for probing the fine structure of transition-metal oxides.
    
    \par The SPT exhibits outstanding accuracy, stability, and generalization in local environment prediction, underscoring its potential for accelerating materials discovery and enabling deeper understanding of structure–property relationships in complex materials.



    \subsection{Performance Analysis of SPT}

    \par It remains an open question whether the SPT method can capture the most significant spectral variations induced by changes in the structure descriptor. To address this, we evaluate its performance in Supplementary Figure 3(S3) against several criteria. First, we examine the quantitative correlation for discrete parameters representing short-range (CN), medium-range (CN2), and long-range (PSGO) structural correlations. Concurrently, we  evaluate the classification capability for key chemical descriptors—oxygen OCN and OS. We then discuss continuous descriptors—NNRS and MOOD—which are local descriptors crucial for experimentalists to understand subtle changes in underlying physical and chemical processes. Figure~S3 illustrates the performance of classification and regression tasks on non-oxygen site datasets, with each dataset encompassing descriptors that reflect distinct dependent characteristics of the local atomic environment. Analyses for oxygen site datasets are detailed in Figure~S4. 
    
    \par For non-oxygen sites, the classification results for the CN, CN2, and PSGO descriptors are shown in Figure S3(a–c). The corresponding confusion matrices and receiver operating characteristic curves show that the model achieves stable, high performance for CN (accuracy and F1-Score $>$ 0.98). A gradual performance reduction is observed for CN2 and PSGO, which correspond to more global structural features. This trend indicates that predictive precision diminishes as the descriptor extends from local coordination to long-range symmetry, due to increased structural complexity and spectral overlap. 

    \par Figure~S3(d-e) additionally presents the model's classification results for the two key chemical descriptors OCN and OS. The model also demonstrated outstanding performance on these two tasks. For OCN, it achieves an F1-Score of 0.983; for OS, the accuracy and F1-Score are 0.969 and 0.969 respectively. These results validate that the SPT method not only captures geometric structural information but also effectively infers chemical features associated with charge states and specific atomic coordination.
    
    \par The regression performance for the continuous descriptors NNRS and MOOD is shown in Figure S3(f–g). The predicted values show a strong linear relationship with the true values, with high coefficients of determination and low RMSE. This confirms the model's effectiveness in capturing quantitative variations in bond-length distortions and oxygen-oxygen spatial correlations. Notably, NNRS, representing the standard deviation of bond lengths, shows near-perfect agreement with the reference line, demonstrating the model's sensitivity to subtle geometric fluctuations. 
    
    \par In summary, the performance trend reveals a clear pattern: predictive accuracy for structural descriptors decreases progressively from CN to PSGO, corresponding to the increasing spatial extent of the structural information encoded in the XANES spectra. This successfully resolves our initial query, demonstrating that while the SPT method is most sensitive to local atomic environments, it remains highly capable of inferring medium to long-range features. Meanwhile, the model's accurate classification of chemical descriptors such as OCN and OS further confirms its high sensitivity to both electronic and chemical environments. The consistently high regression accuracy across all descriptors further validates that the proposed SPT effectively integrates both spectral locality and globality, enabling reliable inference of diverse structural and electronic properties from complex absorption spectra.

\subsection{Synergistic Efficiency: Unifying Structural Predictions via Multi-Target Joint SPT}

    \par To establish a high-fidelity benchmark for XANES characterization, we first implemented the Expert SPT strategy. This approach involves training dedicated models for specific structural descriptors—optimizing weights exclusively for individual metrics such as OS, CN, or NNRS. By isolating the optimization landscape, this strategy minimizes gradient conflict between competing objectives, thereby yielding superior accuracy. However, this specificity incurs a significant computational cost: acquiring a comprehensive structural profile necessitates the sequential training and execution of multiple independent models, which creates substantial redundancy and limits throughput in large-scale screening applications.
    
    \par To resolve this efficiency bottleneck, we propose the Multi-Target Joint SPT framework, a unified architecture capable of simultaneously predicting all structural descriptors from a single spectrum. This framework encompasses a holistic target set: OS, OCN, CN, CN2, PSGO, NNRS, and MOOD for non-oxygen sites; and CN, CN2, PSGO, NNRS, and NNOSM for oxygen sites. The core advantage of this paradigm lies in the construction of a shared feature space, where latent representations generalize across inherently correlated physical tasks. By capturing the underlying chemical bonding and coordination geometry principles, the unified model facilitates efficient cross-task information exchange—transferring inductive biases learned in robust tasks (e.g., CN) to regularize more complex estimations (e.g., NNRS).
    
    \par Figure~\ref{comparion}g and h presents a quantitative assessment of the trade-off between efficiency and accuracy based on the updated benchmarking data. For the non-oxygen dataset (Figure~\ref{comparion}g), the Multi-Target Joint SPT demonstrates remarkable performance parity with the Expert benchmarks. Deviations for foundational classification tasks (OS, OCN, CN, CN2) are contained within a narrow margin of approximately 1.0\%. Interestingly, we observe a counter-intuitive performance gain in complex tasks: the joint model achieves improvements in both PSGO PSGO (rising from 0.517 to 0.523) and MOOD (from 0.684 to 0.690). This suggests that the shared latent representations yield synergistic effects, where simpler tasks provide auxiliary supervision that helps the model navigate the optimization landscape for more abstract geometric descriptors.
    
    \par A similar pattern of synergistic regularization is observed in the oxygen dataset (Figure~\ref{comparion}h). Contrary to the common expectation of task interference in multi-objective learning, the joint model actually enhances performance for challenging structural descriptors. Specifically, the F1-score for PSGO increases from 0.606 to 0.620, and the $R^2$ for NNRS improves marginally from 0.844 to 0.846. While there is a slight trade-off in the highly sensitive NNOSM task ($R^2$ decreases from 0.951 to 0.938), the overall integrity of the predictions remains high. This indicates that the joint framework effectively leverages multidimensional spectral signals to build a more robust, generalized representation of the local atomic environment, rather than overfitting to isolated features.
    
    \par Crucially, these marginal fluctuations in precision are dramatically outweighed by transformative gains in computational efficiency. The Joint SPT framework streamlines the workflow by consolidating up to seven distinct models into a single unified architecture, reducing total training time by approximately 85.0\%. By obviating the need for sequential inference, the model generates an integrated structural profile with minimal latency. This efficiency-accuracy balance is pivotal for high-throughput materials discovery, enabling rapid, multi-dimensional structural characterization without the computational overhead of ensemble expert systems.

\backmatter

\bmhead{Code and Data Availability}
The SPT code is available via GitHub at \url{https://github.com/zsy-suyang/SPT}. The datasets and trained model parameters are publicly available via Hugging Face at \url{https://huggingface.co/datasets/zsysuyang/XAS_SPT} upon the acceptance of the manuscript.

\bmhead{Author Contributions}
F.T. and J.C. designed the study. B.Y.H. performed the first principles calculations to generate the dataset. S.Y.Z. contributed to the methodology development and the model training. F.T., S.Y.Z., B.Y.H., and P.W.X. performed data analysis. All authors contributed to interpreting the results and writing the manuscript.

\bmhead{Acknowledgment}
F.T. acknowledges the National Key R\&D Program of China (Grant No. 2024YFA1210804), National Natural Science Foundation of China (Grant No. 22573085) and a startup fund at Xiamen University. J.C. acknowledges the National Natural Science Foundation of China (Grant Nos.22021001, 22225302, 21991151, 21991150, 92161113, and 20720220009) and the Laboratory of AI for Electrochemistry (AI4EC) and IKKEM (Grant Nos. RD2023100101 andRD2022070501) for financial support.  We acknowledge the Fundamental Research Funds for the Central Universities (AI for Energy Chemistry, Grant No. 20720250005). This work used the computational resources in the IKKEM intelligent computing center.

\bibliography{SPT_bibliography}
\nolinenumbers
\end{document}



\title[Article Title]{Supplementary Information for ``Solving the inverse problem of X-ray absorption spectroscopy via physics-informed deep learning''}

\author[1]{\fnm{Suyang} \sur{Zhong}}
\equalcont{These authors contributed equally to this work.}
\author[1]{\fnm{Boying} \sur{Huang}}
\equalcont{These authors contributed equally to this work.}
\author[2]{\fnm{Pengwei} \sur{Xu}}
\equalcont{These authors contributed equally to this work.}
\author[2]{\fnm{Fanjie} \sur{Xu}}
\author[2]{\fnm{Yuhao} \sur{Zhao}}

\author*[3,2,4]{\fnm{Jun} \sur{Cheng}}\email{chengjun@xmu.edu.cn}
\author*[1,2,4]{\fnm{Fujie} \sur{Tang}}\email{tangfujie@xmu.edu.cn}

\author[5,6,7]{\fnm{Weinan} \sur{E}}
\author[3,4]{\fnm{Zhong-Qun} \sur{Tian}}

\affil[1]{\orgdiv{Pen-Tung Sah Institute of Micro-Nano Science and Technology, Discipline of Intelligent Instrument and Equipment, iChEM}, \orgname{Xiamen University}, \orgaddress{\city{Xiamen}, \postcode{361005}, \country{China}}}

\affil[2]{\orgdiv{Institute of Artificial Intelligence}, \orgname{Xiamen University}, \orgaddress{\city{Xiamen}, \postcode{361005}, \country{China}}}

\affil[3]{\orgdiv{State Key Laboratory of Physical Chemistry of Solid Surfaces, iChEM, College of Chemistry and Chemical Engineering}, \orgname{Xiamen University}, \orgaddress{\city{Xiamen}, \postcode{361005}, \country{China}}}

\affil[4]{\orgdiv{Laboratory of AI for Electrochemistry (AI4EC), Tan Kah Kee Innovation Laboratory (IKKEM)}, \orgaddress{\city{Xiamen}, \postcode{361005}, \country{China}}}

\affil[5]{\orgname{AI for Science Institute}, \orgaddress{\city{Beijing}, \postcode{100080}, \country{China}}}

\affil[6]{\orgdiv{Center for Machine Learning Research}, \orgname{Peking University}, \orgaddress{ \city{Beijing}, \postcode{100871}, \country{China}}}

\affil[7]{\orgdiv{School of Mathematical Sciences}, \orgname{Peking University}, \orgaddress{ \city{Beijing}, \postcode{100871}, \country{China}}}

\maketitle

\section{Data Preparation and Validation}\label{secA1}

    \subsection{Training data generation}
    
    \par FEFF calculations are carried out in real space using \textsc{FEFF}10\cite{kas2021advanced}, with the core-hole screened within the random-phase approximation (\texttt{COREHOLE~RPA}). A self-consistent field (SCF) cluster radius of 7~\AA\ (about 100~atoms) is adopted to ensure convergent muffin-tin potentials, while the full multiple-scattering (FMS) radius is extended to 9~\AA\ to include contributions up to the third coordination shell. The energy grid for XANES is set to 0.04~eV with a Lorentzian broadening of 0.1~eV, and no path-length truncation (\texttt{RPATH~-1}) is applied, guaranteeing sufficient accuracy for the K-edge spectral features. All input files are generated using the \textsc{Lightshow} package\cite{carbone2023lightshow}. The same computational parameters are employed for the spectral data used in the Li-Co-O and amorphous materials tasks.
    
    \subsection{Data Distribution}
    
    \par The K-edge XANES spectra of the following oxide systems are computed: Co--Ni--O, Li--Ni--O, Li--O--Ti, Co--O, Fe--Li--O--P, Co--Li--Mn--Ni--O, Co--Li--Mn--O, Fe--O, Mn--Ni--O, Co--Li--O, Li--Mn--Ni--O, Mn--O, Co--Mn--O, Li--Mn--O, and Ni--O. All crystal structures are acquired from the Materials Project database\cite{jain10materials}, yielding \textbf{56\,565} non-equivalent absorbing site K-edge XANES spectra. The dataset is stratified by element type, with 85\,\% of the spectra reserved for training and the remaining 15\,\% for testing. Experimental reference data are obtained from the LISA XAS Database (\url{https://lisa.iom.cnr.it/xasdb/}) standard samples.
    
    \subsection{Experimental Spectral Alignment}
    

    \par All experimental Co K-edge spectra are first pre-processed following standard background subtraction and normalization protocols~\cite{ravel2005athena, kas2021advanced}. Subsequently, the spectra are truncated to the rising-edge region spanning 7713--7776~eV. To correct the residual systematic offset between theory and experiment, a standard reference foil (Materials Project ID mp-22526, Co K-edge is measured, revealing a constant energy mis-calibration of $\approx$ 0.5~eV. Such systematic deviations are frequently observed in \textit{ab initio} XAS simulations due to the approximations in modeling core-hole potentials or exchange-correlation functionals~\cite{rehr2000theoretical, cabaret2010first}. Consequently, this correction value is uniformly added to every experimental spectrum, bringing the entire data set into register with the computed energy scale. The same two-step alignment procedure---edge-position anchoring followed by a single global shift---can be applied straightforwardly to other absorption edges, ensuring a consistent energy reference for combined computational and experimental studies~\cite{mathew2018high}. 

    \subsection{UMAP for XANES Spectral Embedding}

    To supplement the spectral-structural correlation analysis discussed in the main text, we elaborate in detail on the Uniform Manifold Approximation and Projection (UMAP) algorithm~\cite{mcinnes2018umap}, with particular emphasis on its distinctive efficacy in processing high-dimensional XANES data. The theoretical foundation of UMAP lies in Riemannian geometry and algebraic topology, operating under the assumption that spectral data is distributed across a locally connected Riemannian manifold and possesses a unified fuzzy simplicial representation. Unlike linear dimensionality reduction methods such as Principal Component Analysis (PCA) – which prioritise capturing maximum global variance through orthogonal projection – UMAP constructs a weighted graph within the high-dimensional spectral space. Within this graph, edge weights represent the probability of connections between spectral points. This characteristic proves particularly crucial for XANES data, as its subtle nonlinear variations in pre- or post-edge oscillations encapsulate vital information regarding coordination geometry and electronic transitions~\cite{rehr2000theoretical}.
    
    This algorithm projects these high-dimensional spectral vectors where each energy point constitutes a dimension onto an interpretable two-dimensional manifold (defined by $\mathrm{UMAP}_1$ and $\mathrm{UMAP}_2$) by minimising the cross-entropy between high-dimensional and low-dimensional topological representations. The method's distinctive advantage lies in its ability to balance the preservation of local and global structures. For this dataset, local connectivity ensures spectral clusters originating from approximately identical local environments—such as mild distortions within $M\mathrm{O}_6$ octahedra—are grouped in contiguous regions, while global optimisation preserves macroscopic correlations between distinct chemical species. This dual preservation mechanism effectively filters out random noise, highlighting latent variables governing photon-atom interactions. Consequently, descriptors such as oxidation state (OS) and coordination number (CN) exhibit smooth, continuous gradients. Consequently, distances within the UMAP plane serve as robust surrogate metrics for structural and electronic similarity between distinct atomic sites, confirming that spectral ‘fingerprints’ are not isolated features but continuous nodes within a structured physical space.

\section{SPT Method Configuration}\label{secB1}

    \begin{figure*}[t]
    \begin{center}
        \centering
        \includegraphics[width=\linewidth]{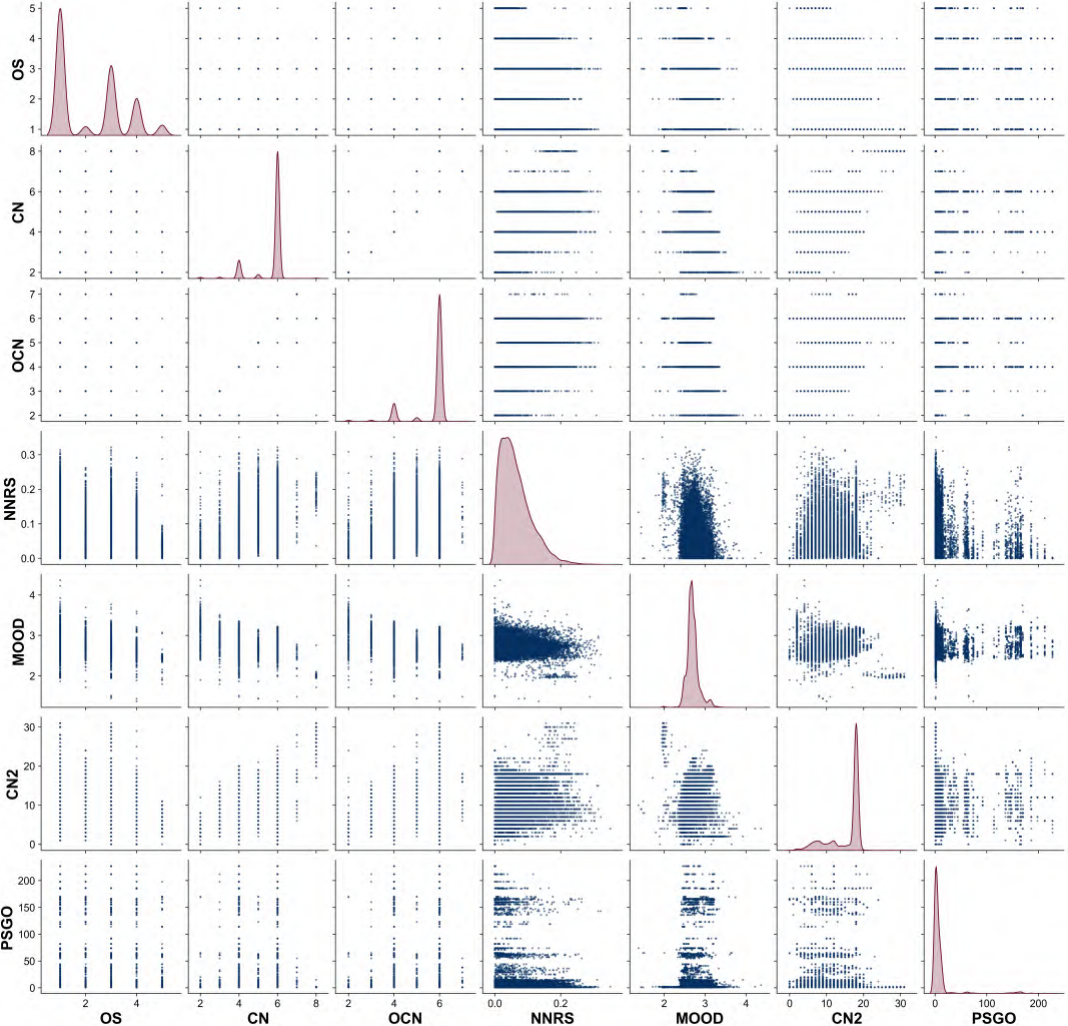}
        \caption{
            \textbf{Pairwise distribution and correlation analysis of structural descriptors at non-oxygen sites.}
            The diagonal subplots display the univariate kernel density estimates for each descriptor, such OS, CN, CN2, PSGO, revealing significant class imbalance. Notably, CN2 and PSGO exhibit characteristic long-tail distributions with high skewness. The off-diagonal scatter plots visualize the joint relationships between descriptor pairs, illustrating the sparse and discrete nature of the raw categorical values, such as the distinct clustering in OS-CN pair. These distributional properties underscore the necessity for category label discretisation.
        }
        \label{non_oxygen_pair}
    \end{center}
    \end{figure*}   

\subsection{Details of model training normalization}
    Prior to model training, a data preprocessing workflow is implemented to ensure numerical stability and statistical validity. Firstly, to safeguard the robustness of training and validation, the dataset is filtered to exclude underrepresented categories. Additionally, labels occurring fewer than three times are also discarded.
    
    Distinct normalisation strategies are applied to spectral energy points and absorption intensity. The energy axis undergoes global normalisation: raw energy values are first compressed via a logarithmic transformation to reduce dynamic range, and then linearly scaled to the unit interval $[0, 1]$ based on the global minimum and maximum energy values within the training set. This ensures standardisation of the input domain across all sample energy grids.
    
    In contrast, spectral intensity employs an intra-sample normalisation scheme. Each XANES spectrum is individually scaled to the range $[0, 1]$ using its own local minimum and maximum intensities as reference points. This sample-based normalisation preserves the relative shape characteristics of absorption edges while eliminating the influence of absolute intensity variations—variations that typically arise from experimental conditions or computational scaling factors.

\subsection{Physicochemical Correlations and Structural Constraints}
    \begin{figure*}[t]
    \begin{center}
        \centering
        \includegraphics[width=\linewidth]{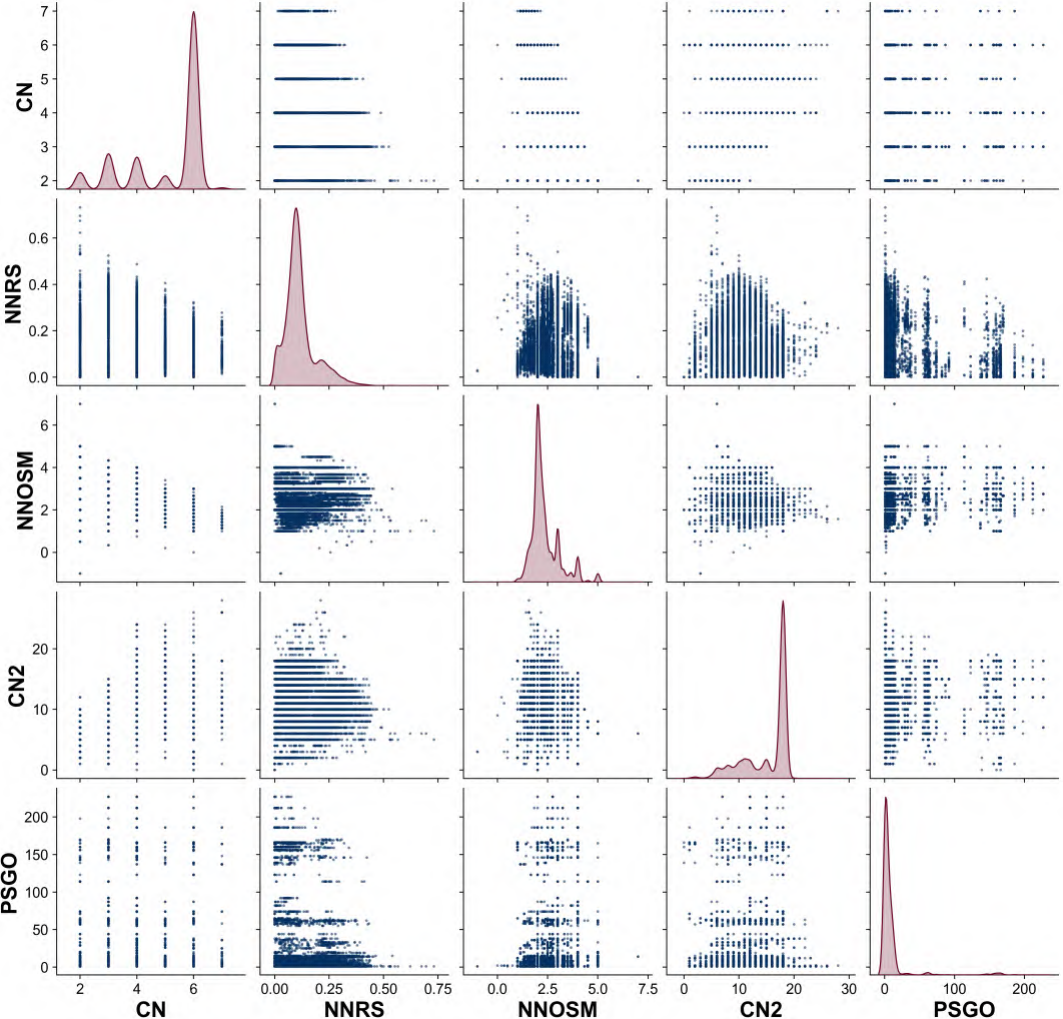}
        \caption{
            \textbf{Pairwise distribution and correlation analysis of structural descriptors at oxygen sites.}
            The diagonal subplots present kernel density estimates for oxygen-specific descriptors, including CN, NNRS, NNOSM, CN2, and PSGO. The off-diagonal scatter plots illustrate significant structural dependencies. The CN–CN2 pairs demonstrate how local anionic coordination governs the broader cationic network density, whereas clustering in NNOSM versus CN reflects charge-neutrality constraints imposed by the surrounding cation species.
        }
        \label{oxygen_pair}
    \end{center}
    \end{figure*}

    Beyond discretisation requirements, the off-diagonal pair distributions in Figure \ref{non_oxygen_pair} reveal intrinsic physicochemical correlations that govern non-oxygen sites.
    
    First, the joint distribution of OS and CN exhibits distinct clustering rather than random dispersion, reflecting fundamental valence-bond and coordination constraints. Specific oxidation states show strong preferences for particular coordination environments, such as concentrated clusters at low oxidation values, implicitly encoding bond-valence principles in which local geometry is strictly constrained by the electronic state of the cation. Regions that lack observed data points within the oxidation-state–coordination-number grid indicate physically unstable or chemically forbidden configurations.
    
    Secondly, a positive topological dependency exists between the CN values of the first coordination sphere and the CN2 values of the second sphere. As illustrated in the CN–CN2 scatter plot, increasing the coordination number of the first layer imposes a lower-bound constraint on the packing density of the second layer. This trend indicates that steric hindrance propagates outward from the central atom, leading to increased density in the secondary coordination shell. The correlations among these descriptors show that, despite the discrete modelling of the target variables, their underlying physical properties exhibit coupled relationships, requiring the model to learn these implicit structural rules.
    
    Parallel structural constraints appear for oxygen sites, as visualised in Figure \ref{oxygen_pair}, although they manifest through anion-specific patterns. The distribution of CN is heavily skewed toward lower integers (predominantly 2, 3, and 4), consistent with the typical coordination geometries of oxide anions. The correlation between CN and CN2 at oxygen sites mirrors the topological propagation observed at cationic sites, where specific local coordination environments (CN) determine the accessible range of extended-network connectivity (CN2). Furthermore, the joint distributions involving NNRS and NNOSM display non-uniform and well-defined clustering patterns. These patterns suggest that the local chemical environment of oxygen is governed by strict charge-balancing rules imposed by the connected cations, producing discrete “islands of stability’’ in the descriptor space rather than a continuous manifold.

\subsection{Feature Embedding Method}

    Spectral preprocessing in SPT method requires input features to undergo tokenisation and embedding processing. For different feature types, we employ differentiated strategies: categorical features can be processed via standard embedding lookup; whereas continuous numerical features necessitate specialised encoding schemes to effectively project them onto the $d$-dimensional embedding space. To determine the optimal approach, we conducted a comparative analysis of three distinct numerical feature embedding architectures.
    
    \begin{enumerate}
        \item \textbf{Linear transformation:} Performs feature-level linear affine transformations on each numerical input. This method maximises the preservation of the original numerical features' structural integrity during dimensionality reduction, ensuring that relationships such as relative band positions and energy distribution trends in spectral data remain intact.
        
        \item \textbf{MLP-based Embedder:} Processes each numerical feature through a dedicated multi-layer perceptron, enabling the model to independently learn complex nonlinear transformations for each feature, thereby significantly enhancing representational capacity.
        
        \item \textbf{Transformer Embedder:} Collectively processes numerical features via self-attention mechanisms to capture latent interactions between numerical features.
    \end{enumerate}
    
    We integrate these three embedding methods into the core SPT method and experimentally evaluate them on structural descriptor prediction tasks. Surprisingly, during the feature embedding stage, simple linear transformations generally outperform more complex multi-layer perceptrons and Transformer encoders while demonstrating higher computational efficiency.
    
    We contend that interpretable transformations that avoid non-linear couplings prevent excessive distortion of signal structure during early feature extraction, thereby providing subsequent models with more stable and physically coherent input representations.
    
    Consequently, introducing nonlinear transformations or self-attention mechanisms during feature embedding for specific tasks and data distributions may prove not only unbeneficial but potentially detrimental to generalisation capabilities and optimisation outcomes. Given the advantages of linear transformations in terms of performance, computational efficiency, and structural simplicity, we adopt them as the standard approach for numerical embedding within our framework.

\section{Additional Experimental Details}\label{secC1}

\subsection{Evaluation Criteria}

    \par To quantitatively assess the performance of the SPT method in both classification and regression tasks, we employ a set of standard evaluation metrics widely used in machine learning. These metrics enable consistent and interpretable comparison across descriptors of different spatial ranges and tasks of distinct objectives.
    
    \par For the classification tasks, we use Precision, Recall, Accuracy, and F1-Score to evaluate the prediction quality across all classes. Let $TP$, $TN$, $FP$, and $FN$ denote the number of true positives, true negatives, false positives, and false negatives, respectively. The metrics are defined as follows:
    
    \begin{equation}
    \text{Precision} = \frac{TP}{TP + FP}
    \end{equation}
    
    \begin{equation}
    \text{Recall} = \frac{TP}{TP + FN}
    \end{equation}
    
    \begin{equation}
    \text{Accuracy} = \frac{TP + TN}{TP + TN + FP + FN}
    \end{equation}
    
    \begin{equation}
    \text{F1-Score} = 2 \times \frac{\text{Precision} \times \text{Recall}}{\text{Precision} + \text{Recall}}
    \end{equation}
    
    \par In the field of spectral classification, these metrics collectively evaluate a model's ability to distinguish structural environments from spectral fine-structure. Precision reflects the reliability of predicted categories, recall assesses the completeness of correctly detected instances, while the F1-Score provides a coordinated balance between the two. High accuracy and F1-Score indicate that the SPT method effectively captures discriminative spectral patterns corresponding to local coordination features.

    \par For the regression tasks, which aim to predict continuous geometric and electronic descriptors such as NNRS and NNOSM, we adopt the Mean Squared Error (MSE), Mean Absolute Error (MAE), Root Mean Squared Error (RMSE), and the coefficient of determination ($R^2$). Let $y_i$ and $\hat{y}_i$ denote the true and predicted values for the $i$-th sample, and $\bar{y}$ the mean of all true values. The metrics are expressed as:
    
    \begin{equation}
    \text{MSE} = \frac{1}{n}\sum_{i=1}^{n}(y_i - \hat{y}_i)^2
    \end{equation}
    
    \begin{equation}
    \text{MAE} = \frac{1}{n}\sum_{i=1}^{n}|y_i - \hat{y}_i|
    \end{equation}
    
    \begin{equation}
    \text{RMSE} = \sqrt{\frac{1}{n}\sum_{i=1}^{n}(y_i - \hat{y}_i)^2}
    \end{equation}
    
    \begin{equation}
    R^2 = 1 - \frac{\sum_{i=1}^{n}(y_i - \hat{y}_i)^2}{\sum_{i=1}^{n}(y_i - \bar{y})^2}
    \end{equation}

    \par The evaluation metrics collectively form a rigorous assessment framework for examining the SPT capability to infer discrete and continuous structural features from spectral data, thereby ensuring the comparability and interpretability of all performance analysis results within the work.

\subsection{Non-oxygen Site Model Performance Comparison}

    \par For the non-oxygen site dataset, Table~\ref{tab:descriptor_non_oxy_cls} summarises the classification performance of different models across five descriptor types. Traditional ensemble methods demonstrate stable performance~\cite{breiman2001random,chen2016xgboost}, whilst Transformer-based models  exhibit stronger generalisation capabilities~\cite{bentsen2023spatio, cheng2024arithmetic}. Notably, our proposed SPT method achieves the best results across all descriptor types, fully demonstrating its capability to capture spectral dependencies in both the frequency domain and graph structure.
    
    \begin{table*}[htbp]
        \centering
        \caption{Performance comparison of classification tasks across non-oxygen site descriptors (values in \%). The best results are highlighted in \textbf{bold}, and the second-best results are \underline{underlined}.}
        \label{tab:descriptor_non_oxy_cls}
        \resizebox{\textwidth}{!}{
        \begin{tabular}{lcccccccc}
        \hline
        \textbf{Descriptor} & \textbf{Metric} & \textbf{Decision Tree} & \textbf{Extra Trees} & \textbf{Random Forests} & \textbf{XGBoost} & \textbf{FTTransformer} & \textbf{AMFormer} & \textbf{SPT(ours)} \\
        \hline
        \multirow{4}{*}{\textbf{OS}} 
        & Precision & 94.28 & 94.53 & \underline{96.57} & 96.49 & 95.71 & 96.21 & \textbf{96.93} \\
        & Recall    & 94.32 & 94.54 & \underline{96.59} & 96.52 & 95.74 & 96.22 & \textbf{96.95} \\
        & ACC       & 94.32 & 94.54 & \underline{96.59} & 96.52 & 95.74 & 96.22 & \textbf{96.95} \\
        & F1        & 94.30 & 94.52 & \underline{96.55} & 96.49 & 95.72 & 96.21 & \textbf{96.92} \\
        \hline
        \multirow{4}{*}{\textbf{CN}} 
        & Precision & 95.83 & 95.49 & 97.20 & 97.65 & 97.54 & \underline{97.84} & \textbf{98.15} \\
        & Recall    & 95.59 & 95.42 & 97.42 & 97.81 & 97.59 & \underline{97.88} & \textbf{98.17} \\
        & ACC       & 95.59 & 95.42 & 97.42 & 97.81 & 97.59 & \underline{97.88} & \textbf{98.17} \\
        & F1        & 95.69 & 95.45 & 97.08 & 97.67 & 97.55 & \underline{97.85} & \textbf{98.14} \\
        \hline
        \multirow{4}{*}{\textbf{OCN}} 
        & Precision & 95.86 & 95.71 & 97.57 & 97.66 & \underline{97.84} & 97.74 & \textbf{98.28} \\
        & Recall    & 95.86 & 95.57 & 97.61 & 97.83 & \underline{97.90} & 97.78 & \textbf{98.32} \\
        & ACC       & 95.86 & 95.57 & 97.61 & 97.83 & \underline{97.90} & 97.78 & \textbf{98.32} \\
        & F1        & 95.85 & 95.63 & 97.29 & 97.65 & \underline{97.87} & 97.75 & \textbf{98.29} \\
        \hline
        \multirow{4}{*}{\textbf{CN2}} 
        & Precision & 75.61 & 74.28 & \underline{78.97} & 77.93 & 76.83 & 78.38 & \textbf{80.75} \\
        & Recall    & 74.93 & 73.93 & \underline{81.46} & 79.95 & 77.81 & 79.73 & \textbf{81.63} \\
        & ACC       & 74.93 & 73.93 & \underline{81.24} & 79.95 & 77.81 & 79.73 & \textbf{81.63} \\
        & F1        & 75.21 & 74.06 & \underline{79.52} & 78.59 & 77.20 & 78.92 & \textbf{81.10} \\
        \hline
        \multirow{4}{*}{\textbf{PSGO}} 
        & Precision & 40.79 & 37.82 & \underline{50.09} & 46.29 & 41.64 & 43.70 & \textbf{50.66} \\
        & Recall    & 40.80 & 37.90 & \underline{53.18} & 51.16 & 45.90 & 47.28 & \textbf{53.50} \\
        & ACC       & 40.80 & 37.90 & \underline{53.18} & 51.16 & 45.90 & 47.28 & \textbf{53.50} \\
        & F1        & 40.76 & 37.83 & \underline{47.75} & 46.92 & 42.55 & 44.65 & \textbf{51.69} \\
        \hline
        \end{tabular}
        }
    \end{table*}

    \par For regression-type descriptors in non-oxygen site datasets, Table~\ref{tab:descriptor_non_oxy_reg} presents results for the NNRS and MOOD tasks. Our proposed SPT method achieves the lowest MSE and MAE across both tasks while attaining the highest $R^2$ values, highlighting its remarkable capability to learn smooth structure-spectrum mappings in the continuous domain.
    
    \begin{table*}[htbp]
        \centering
        \caption{Performance comparison of regression tasks across non-oxygen site descriptors. MSE, MAE, and RMSE are in decimal notation; $R^2$ values are in \%. Note: For MSE, MAE, and RMSE, lower values are better; for $R^2$, higher values are better.}
        \label{tab:descriptor_non_oxy_reg}
        \resizebox{\textwidth}{!}{
        \begin{tabular}{lcccccccc}
        \hline
        \textbf{Descriptor} & \textbf{Metric} & \textbf{Decision Tree} & \textbf{Extra Trees} & \textbf{Random Forests} & \textbf{XGBoost} & \textbf{FTTransformer} & \textbf{AMFormer} & \textbf{SPT(ours)} \\
        \hline
        \multirow{4}{*}{\textbf{NNRS}} 
        & MSE   & 0.0014 & 0.0013 & \underline{0.0006} & 0.0007 & 0.0009 & \underline{0.0006} & \textbf{0.0005} \\
        & MAE   & 0.0223 & 0.0218 & 0.0164 & 0.0171 & 0.0207 & \underline{0.0152} & \textbf{0.0128} \\
        & RMSE  & 0.0370 & 0.0362 & 0.0253 & 0.0258 & 0.0299 & \underline{0.0252} & \textbf{0.0224} \\
        & $R^2$ & 35.79 & 38.45 & 70.05 & 68.93 & 58.15 & \underline{70.28} & \textbf{76.44} \\
        \hline
        \multirow{4}{*}{\textbf{MOOD}} 
        & MSE   & 0.0155 & 0.0161 & \underline{0.0076} & 0.0084 & 0.0107 & 0.0201 & \textbf{0.0070} \\
        & MAE   & 0.0692 & 0.0696 & \underline{0.0516} & 0.0559 & 0.0676 & 0.0969 & \textbf{0.0467} \\
        & RMSE  & 0.1243 & 0.1269 & \underline{0.0869} & 0.0918 & 0.1037 & 0.1419 & \textbf{0.0834} \\
        & $R^2$ & 29.85 & 26.87 & \underline{65.71} & 61.72 & 51.21 & 8.68 & \textbf{68.42} \\
        \hline
        \end{tabular}
        }
    \end{table*}

\subsection{Oxygen Site Model Performance Comparison}
    \par For the oxygen site dataset, the SPT method demonstrates comprehensive performance advantages. For classification tasks, regardless of whether CN, CN2, or PSGO descriptors are employed, Table~\ref{tab:descriptor_oxygen_cls} shows that SPT achieves significantly superior precision, recall, accuracy, and F1-Score compared to traditional machine learning models, as well as deep learning baseline models. 
    
    \par In regression tasks, Table~\ref{tab:descriptor_oxygen_reg} shows that SPT similarly excels with NNRS and NNOSM features, achieving the lowest error metrics  and the highest coefficient of determination $R^2$ values. Notably, in the complex NNOSM prediction task, the SPT attains an $R^2$ value of 95.08\%, significantly surpassing other models. Its error metrics (MSE 0.0236, MAE 0.0712, RMSE 0.1538) also markedly outperformed baseline models.
    
    \par Therefore, the SPT comprehensively outperforms existing methods in both classification and regression tasks, demonstrating enhanced predictive capabilities particularly in complex scenarios.
    
    \begin{table*}[htbp]
        \centering
        \caption{Performance comparison of classification models across different descriptors for the oxygen site dataset.}
        \label{tab:descriptor_oxygen_cls}
        \resizebox{\textwidth}{!}{
        \begin{tabular}{lcccccccc}
        \hline
        \textbf{Descriptor} & \textbf{Metric} & \textbf{Decision Tree} & \textbf{Extra Trees} & \textbf{Random Forests} & \textbf{XGBoost} & \textbf{FTTransformer} & \textbf{AMFormer} & \textbf{SPT(ours)} \\
        \hline
        \multirow{4}{*}{\textbf{CN}} 
        & Precision & 89.90 & 87.16 & 93.48 & 93.65 & \underline{94.67} & 94.62 & \textbf{95.16} \\
        & Recall    & 89.75 & 87.08 & 93.45 & 93.61 & 94.52 & \underline{94.54} & \textbf{95.14} \\
        & ACC       & 89.75 & 87.08 & 93.44 & 93.61 & 94.52 & \underline{94.54} & \textbf{95.14} \\
        & F1        & 89.80 & 87.10 & 93.37 & 93.56 & 94.52 & \underline{94.55} & \textbf{95.12} \\
        \hline 
        \multirow{4}{*}{\textbf{CN2}} 
        & Precision & 73.87 & 71.36 & 78.57 & 80.47 & 80.57 & \underline{80.76} & \textbf{82.47} \\
        & Recall    & 73.42 & 70.87 & 80.14 & \underline{81.46} & 81.03 & 80.80 & \textbf{83.11} \\
        & ACC       & 73.42 & 70.87 & 80.14 & \underline{81.46} & 81.03 & 80.80 & \textbf{83.11} \\
        & F1        & 73.59 & 71.08 & 78.81 & 80.67 & \underline{80.74} & \underline{80.74} & \textbf{82.70} \\
        \hline
        \multirow{4}{*}{\textbf{PSGO}} 
        & Precision & 45.59 & 40.19 & 54.68 & 52.43 & 54.52 & \underline{56.42} & \textbf{59.92} \\
        & Recall    & 45.20 & 40.42 & 57.34 & 56.79 & 57.34 & \underline{58.64} & \textbf{62.13} \\
        & ACC       & 45.20 & 40.42 & 57.34 & 56.79 & 57.34 & \underline{58.64} & \textbf{62.13} \\
        & F1        & 45.34 & 40.26 & 51.31 & 52.29 & 55.35 & \underline{57.19} & \textbf{60.55} \\
        \hline
        \end{tabular}
        }
    \end{table*}
    
    \begin{table*}[htbp]
        \centering
        \caption{Performance comparison of regression models across different descriptors for the oxygen site dataset.}
        \label{tab:descriptor_oxygen_reg}
        \resizebox{\textwidth}{!}{
        \begin{tabular}{lcccccccc}
        \hline
        \textbf{Descriptor} & \textbf{Metric} & \textbf{Decision Tree} & \textbf{Extra Trees} & \textbf{Random Forests} & \textbf{XGBoost} & \textbf{FTTransformer} & \textbf{AMFormer} & \textbf{SPT(ours)} \\
        \hline 
        \multirow{4}{*}{\textbf{NNRS}} 
        & MSE   & 0.0021 & 0.0020 & 0.0010 & 0.0012 & 0.0010 & \underline{0.0009} & \textbf{0.0008} \\
        & MAE   & 0.0268 & 0.0263 & 0.0192 & 0.0209 & 0.0205 & \underline{0.0184} & \textbf{0.0155} \\
        & RMSE  & 0.0461 & 0.0447 & 0.0316 & 0.0343 & 0.0324 & \underline{0.0305} & \textbf{0.0283} \\
        & $R^2$ & 58.34 & 60.86 & 80.42 & 77.00 & 79.50 & \underline{81.78} & \textbf{84.37} \\
        \hline
        \multirow{4}{*}{\textbf{NNOSM}} 
        & MSE   & 0.1025 & 0.1004 & 0.0449 & 0.0461 & \underline{0.0315} & 0.0452 & \textbf{0.0236} \\
        & MAE   & 0.1444 & 0.1455 & 0.1148 & 0.1294 & \underline{0.1100} & 0.1512 & \textbf{0.0712} \\
        & RMSE  & 0.3202 & 0.3168 & 0.2120 & 0.2147 & \underline{0.1774} & 0.2126 & \textbf{0.1538} \\
        & $R^2$ & 78.69 & 79.14 & 90.66 & 90.42 & \underline{93.46} & 90.60 & \textbf{95.08} \\
        \hline
        \end{tabular}
        }
    \end{table*}

\subsection{Oxygen Site Performance Analysis of SPT}

    \begin{figure*}[t]
    \centering
    \includegraphics[width=\linewidth]{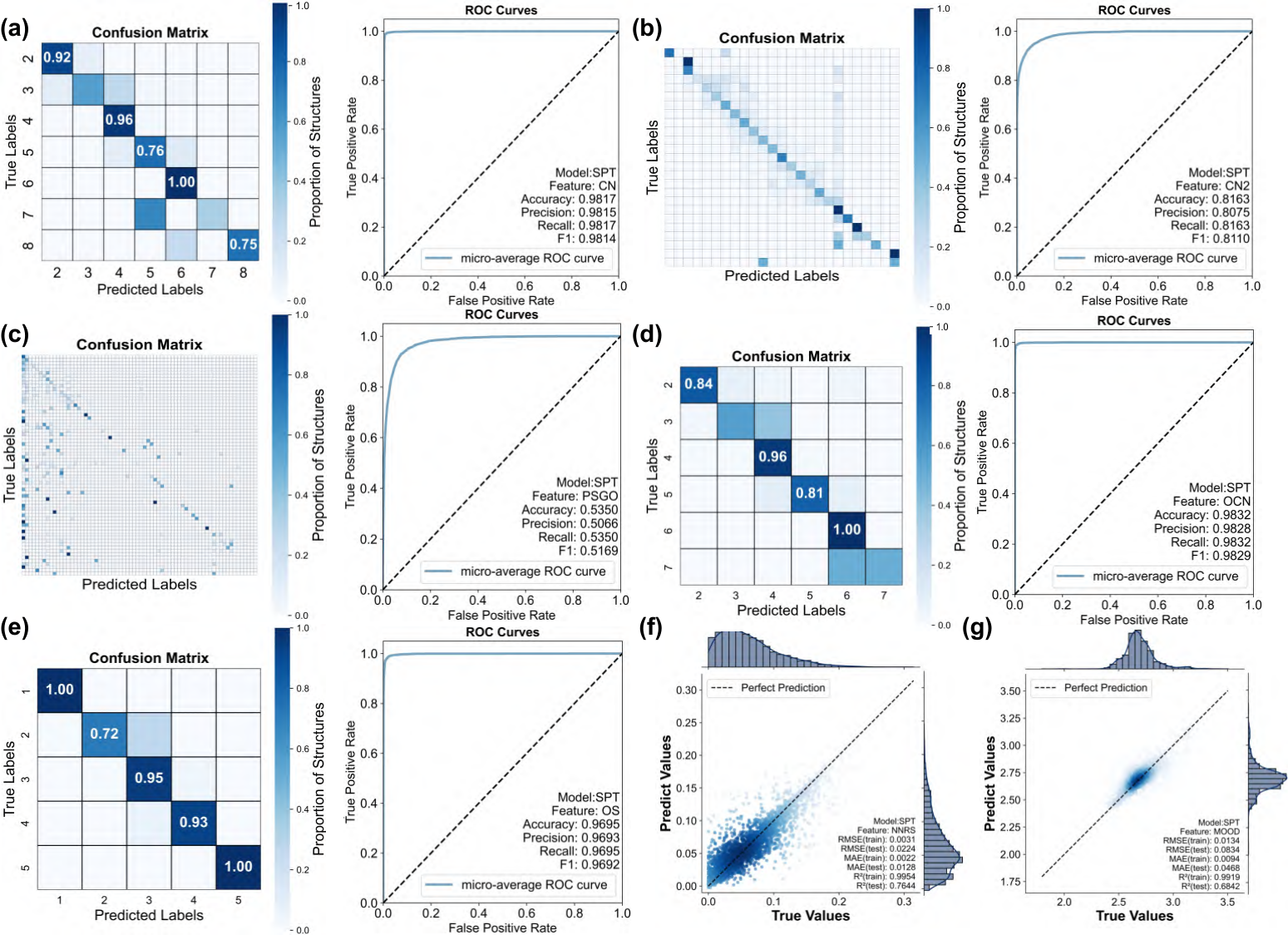}
    \caption{
        \textbf{Detailed evaluation of SPT on non-oxygen structural descriptors.}
        (a–c) Classification performance metrics with increasing descriptor space coverage: short-range CN, medium-range CN2, and long-range PSGO.
        (d, e) Classification results for chemical descriptors OCN and OS.
        (f, g) Regression performance of continuous descriptors NNRS and MOOD.
    }
    \label{metal_ROC}
    \end{figure*}

    \begin{figure*}[h] 
    \centering 
    \includegraphics[width=\linewidth]{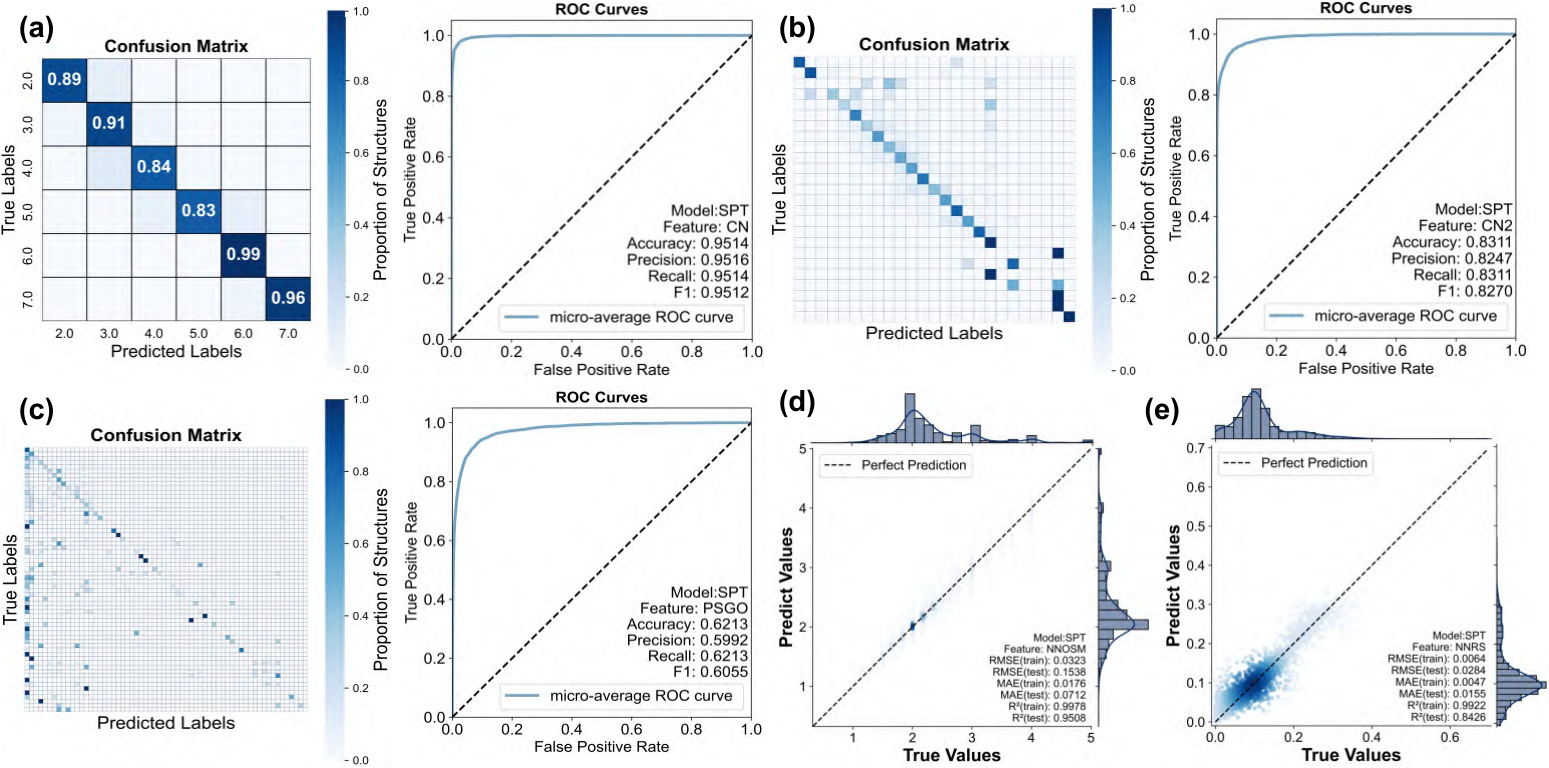} 
        \caption{
        \textbf{Performance evaluation of the SPT framework on oxygen site descriptors.} 
        (a–c) Classification metrics for structural descriptors with increasing spatial scope: short-range (CN), mid-range (CN2), and long-range (PSGO). 
        (d–e) Regression performance for continuous electronic and geometric descriptors (NNOSM and NNRS). The results demonstrate high predictive fidelity for local features, consistent with non-oxygen site analysis.
    } 
    \label{oxygen_ROC} 
    \end{figure*}

    \par To assess the generalizability of the SPT framework, we extended our evaluation to the oxygen site dataset (Figure \ref{oxygen_ROC}). The results corroborate the findings presented in the main text, exhibiting performance trends that strictly mirror those observed for non-oxygen sites.
    
    \par In classification tasks (Figure \ref{oxygen_ROC} a–c), the model demonstrates a clear hierarchy of predictive fidelity governed by spatial locality. For the short-range Coordination Number (CN), the SPT achieves superior performance (Accuracy: 0.9514, F1-Score: 0.9512). The pronounced diagonal dominance in the confusion matrix confirms that the spectral fine structure effectively encodes the first coordination shell. However, predictive accuracy attenuates as the descriptor scope expands to the medium-range (CN2) and long-range (PSGO). Specifically, the F1-Score decreases to 0.8270 for CN2 and further to 0.6055 for PSGO. This gradient in performance recapitulates the intrinsic physics of XAS, where photoelectron scattering signals decay with distance, thereby increasing the difficulty of retrieving long-range structural information.
    
    \par For regression tasks (Figure \ref{oxygen_ROC} d–e), the model exhibits robust capabilities in quantifying electronic and geometric variations. The prediction of the Nearest Neighbor Oxidation State Mean (NNOSM) yields exceptional precision, with an $R^2$ of 0.9508 and an RMSE of 0.1538, evidenced by the tight clustering of data points along the identity line. Similarly, the regression for the Nearest Neighbor Bond Length (NNRS) achieves an $R^2$ of 0.8437. Although the $R^2$ is marginally lower than that of NNOSM, the extremely low RMSE (0.0283) and MAE (0.0155) underscore the model's sensitivity to subtle geometric fluctuations in the local oxygen environment.
    
    \par In summary, the analysis of oxygen sites serves as a robust validation of the SPT framework. By accurately inferring electronic structure features and maintaining high sensitivity to local geometric environments, the SPT demonstrates significant potential as a universal tool for resolving structural and electronic properties across diverse chemical systems.

\subsection{Comparative Robustness Analysis against Baseline Models}

    To strictly evaluate the stability of the SPT framework under stochastic spectral distortions, we incorporated a comparative robustness analysis against the Random Forest (RF) algorithm. RF was selected as the comparative baseline because it demonstrated the second-best overall performance among all conventional machine learning estimators in the noise-free benchmarking (refer to Main Text Figure~\ref{comparion}). Independent Gaussian noise with intensities ranging from $\sigma \in [0.05, 0.30]$ was injected into the test spectra to simulate the signal degradation typical of operando experimental environments.
    
    Supplementary Table~\ref{tab:noise_robustness_non_oxy} details the performance trajectories for non-oxygen site descriptors. In classification tasks, SPT demonstrates superior stability across most metrics, particularly for the OS descriptor, where it maintains an F1-score of 0.9102 at maximum noise ($\sigma=0.30$), significantly surpassing RF (0.8148). This indicates that SPT has successfully encoded the global spectral fingerprints of oxidation states, which remain distinguishable even amidst severe fluctuations. It is acknowledged that under extreme noise conditions ($\sigma \geq 0.20$), RF exhibits marginally higher stability in specific local descriptors (NNRS). This phenomenon can be attributed to the discrete, threshold-based nature of decision trees, which effectively acts as a low-pass filter against continuous spectral jitter. In contrast, the SPT, operating as a continuous function approximator, prioritizes fitting the precise spectral morphology, resulting in a slight sensitivity trade-off to maintain its high precision in low-noise regimes.
    
    The scenario differs significantly for oxygen site descriptors, as presented in Supplementary Table~\ref{tab:noise_robustness_oxygen}, where SPT establishes comprehensive dominance. Unlike the non-oxygen case, SPT outperforms RF across all noise levels and descriptor categories. For the CN2 descriptor, SPT maintains a robust F1-score of 0.7715 at $\sigma=0.30$, whereas RF declines to 0.7428. Most notably, in regression tasks (NNOSM and NNRS), the deep learning architecture of SPT proves far more effective than the ensemble approach. For instance, in the NNOSM task under severe noise ($\sigma=0.30$), SPT retains an $R^2$ of 0.9017, widely outstripping RF (0.7326). This suggests that for the complex, highly coupled spectral features typical of oxygen environments, the deep feature extraction capability of SPT is essential for disentangling genuine structural signals from background fluctuations, validating its practical utility for processing raw experimental data.

    \begin{table*}[htbp]
        \centering
        \caption{\textbf{Robustness analysis of non-oxygen site descriptors under varying Gaussian noise levels.} The table compares the performance of SPT (ours) against the Random Forest (RF) baseline. Classification performance is measured by F1-Score, and regression performance by $R^2$. Best results are highlighted in \textbf{bold}.}
        \label{tab:noise_robustness_non_oxy}
        \resizebox{\textwidth}{!}{
        \begin{tabular}{llcccccc}
        \hline
        \multirow{2}{*}{\textbf{Descriptor}} & \multirow{2}{*}{\textbf{Model}} & \multicolumn{6}{c}{\textbf{Noise Level ($\pm \sigma$)}} \\
         & & \textbf{0.05} & \textbf{0.10} & \textbf{0.15} & \textbf{0.20} & \textbf{0.25} & \textbf{0.30} \\
        \hline
        \multicolumn{8}{c}{\textit{Classification (F1-Score)}} \\
        \hline
        \multirow{2}{*}{\textbf{OS}} 
        & RF & 0.9278 & 0.8751 & 0.8089 & 0.8147 & 0.8042 & 0.8148 \\
        & SPT (ours) & \textbf{0.9536} & \textbf{0.9455} & \textbf{0.9340} & \textbf{0.9287} & \textbf{0.9226} & \textbf{0.9102} \\
        \hline
        \multirow{2}{*}{\textbf{CN}} 
        & RF & 0.9675 & 0.9557 & 0.9520 & 0.9491 & 0.9440 & \textbf{0.9425} \\
        & SPT (ours) & \textbf{0.9816} & \textbf{0.9769} & \textbf{0.9682} & \textbf{0.9644} & \textbf{0.9587} & 0.9358 \\
        \hline
        \multirow{2}{*}{\textbf{OCN}} 
        & RF & 0.9569 & 0.9394 & 0.9407 & 0.9334 & 0.9361 & 0.9363 \\
        & SPT (ours) & \textbf{0.9807} & \textbf{0.9762} & \textbf{0.9678} & \textbf{0.9634} & \textbf{0.9539} & \textbf{0.9375} \\
        \hline
        \multirow{2}{*}{\textbf{CN2}} 
        & RF & \textbf{0.7590} & \textbf{0.7339} & 0.7100 & 0.6855 & 0.7043 & 0.6436 \\
        & SPT (ours) & 0.7576 & 0.7297 & \textbf{0.7286} & \textbf{0.7304} & \textbf{0.7240} & \textbf{0.7155} \\
        \hline
        \multirow{2}{*}{\textbf{PSGO}} 
        & RF & 0.3514 & 0.2806 & 0.2713 & 0.2753 & 0.2760 & 0.2431 \\
        & SPT (ours) & \textbf{0.4673} & \textbf{0.4309} & \textbf{0.4191} & \textbf{0.4022} & \textbf{0.3914} & \textbf{0.3812} \\
        \hline
        \multicolumn{8}{c}{\textit{Regression ($R^2$)}} \\
        \hline
        \multirow{2}{*}{\textbf{NNRS}} 
        & RF & 0.5735 & 0.4946 & \textbf{0.4613} & \textbf{0.4257} & \textbf{0.3956} & \textbf{0.3679} \\
        & SPT (ours) & \textbf{0.6615} & \textbf{0.5329} & 0.4380 & 0.3678 & 0.3247 & 0.3325 \\
        \hline
        \multirow{2}{*}{\textbf{MOOD}} 
        & RF & 0.6178 & 0.5593 & 0.5078 & 0.4606 & 0.4365 & 0.4055 \\
        & SPT (ours) & \textbf{0.6679} & \textbf{0.6383} & \textbf{0.5908} & \textbf{0.5302} & \textbf{0.4731} & \textbf{0.4587} \\
        \hline
        \end{tabular}
        }
    \end{table*}
    
    \begin{table*}[htbp]
        \centering
        \caption{\textbf{Robustness analysis of oxygen site descriptors under varying Gaussian noise levels.} Comparisons are made between SPT (ours) and the Random Forest (RF) baseline. Best results are highlighted in \textbf{bold}.}
        \label{tab:noise_robustness_oxygen}
        \resizebox{\textwidth}{!}{
        \begin{tabular}{llcccccc}
        \hline
        \multirow{2}{*}{\textbf{Descriptor}} & \multirow{2}{*}{\textbf{Model}} & \multicolumn{6}{c}{\textbf{Noise Level ($\pm \sigma$)}} \\
         & & \textbf{0.05} & \textbf{0.10} & \textbf{0.15} & \textbf{0.20} & \textbf{0.25} & \textbf{0.30} \\
        \hline
        \multicolumn{8}{c}{\textit{Classification (F1-Score)}} \\
        \hline
        \multirow{2}{*}{\textbf{CN}} 
        & RF & 0.9305 & 0.9241 & 0.9191 & 0.9076 & 0.9015 & 0.8859 \\
        & SPT (ours) & \textbf{0.9507} & \textbf{0.9472} & \textbf{0.9410} & \textbf{0.9305} & \textbf{0.9204} & \textbf{0.9065} \\
        \hline

        \multirow{2}{*}{\textbf{CN2}} 
        & RF & 0.7855 & 0.7788 & 0.7677 & 0.7608 & 0.7503 & 0.7428 \\
        & SPT (ours) & \textbf{0.8243} & \textbf{0.8178} & \textbf{0.8142} & \textbf{0.8068} & \textbf{0.7921} & \textbf{0.7715} \\
        \hline
        \multirow{2}{*}{\textbf{PSGO}} 
        & RF & 0.4536 & 0.4125 & 0.3687 & 0.3457 & 0.3221 & 0.3060 \\
        & SPT (ours) & \textbf{0.6012} & \textbf{0.5920} & \textbf{0.5739} & \textbf{0.5540} & \textbf{0.5330} & \textbf{0.5088} \\
        \hline
        \multicolumn{8}{c}{\textit{Regression ($R^2$)}} \\
        \hline
        \multirow{2}{*}{\textbf{NNRS}} 
        & RF & 0.7931 & 0.7821 & 0.7683 & 0.7542 & 0.7398 & 0.7271 \\
        & SPT (ours) & \textbf{0.8438} & \textbf{0.8404} & \textbf{0.8327} & \textbf{0.8225} & \textbf{0.8101} & \textbf{0.7968} \\
        \hline
        \multirow{2}{*}{\textbf{NNOSM}} 
        & RF & 0.8340 & 0.8031 & 0.7774 & 0.7607 & 0.7464 & 0.7326 \\
        & SPT (ours) & \textbf{0.9496} & \textbf{0.9477} & \textbf{0.9425} & \textbf{0.9337} & \textbf{0.9214} & \textbf{0.9017} \\
        \hline
        \end{tabular}
        }
    \end{table*}

\newpage
\bibliographystyle{unsrt}
\bibliography{SPT_SI-bibliography}
\nolinenumbers